\documentclass[aps,prx,twocolumn,groupedaddress,floatfix,sort&compress,longnamesfirst]{revtex4-2}%revtex4-2

\usepackage{amsmath,amssymb}
\usepackage{bm,bbm}
\usepackage{hyperref}
\hypersetup{
	colorlinks=true,
	linkcolor=blue,
	filecolor=blue,
	urlcolor=blue,
	citecolor=blue,
}
\usepackage{graphicx}% Include figure files
\graphicspath{{figure/}}

\begin{document}

% Use the 'preprintnumbers' class option to override journal defaultts
%\preprint{}

\title{Quantum Signatures of Topological Phase in Bosonic Quadratic System}


%\thanks{}
\author{Yaohua Li$^{1}$}
\author{Yong-Chun Liu$^{1, 2}$}
% \author{coldatom}
%\thanks{}
\email{ycliu@tsinghua.edu.cn}
\affiliation{$^{1}$State Key Laboratory of Low-Dimensional Quantum Physics, Department of Physics, Tsinghua University, Beijing 100084, P. R. China}
\affiliation{$^{2}$Frontier
Science Center for Quantum Information, Beijing 100084, China}


\date{\today}

\begin{abstract}
	Quantum entanglement and classical topology are two distinct phenomena that are difficult to be connected together. Here we discover that an open bosonic quadratic chain exhibits topology-induced entanglement effect.
When the system is in the topological phase, the edge modes can be entangled in the steady state, while no entanglement appears in the trivial phase.
%We find two edge modes can be entangled after a dissipative evolution when the system is in the topological phase.
%From a two-mode system, we show the stationary entanglement originates from the coupling to the environment fluctuation and will be greatly suppressed if such a coupling (denoted by the dissipation rate) is much smaller than the intrasystem couplings.
This finding is verified through the covariance approach based on the quantum master equations, which provide exact numerical results without truncation process. We also obtain concise approximate analytical results through the quantum Langevin equations, which perfectly agree with the exact numerical results.
We show the topological edge states exhibit near-zero eigenenergies located in the band gap and are separated from the bulk eigenenergies, which match the system-environment coupling (denoted by the dissipation rate) and thus the squeezing correlations can be enhanced.
%From a two-mode system, we show the stationary entanglement originates from the coupling to the environment fluctuation and will be greatly suppressed if such a coupling (denoted by the dissipation rate) is much smaller than the intrasystem couplings.
%The near-zero eigenvalues of topological edge states can greatly enhance the squeezing correlations that are suppressed by the small dissipation rate, especially between two edge modes.
%By approximately solving the Langevin equations, we show the squeezing correlations reach a maximum when the absolute eigenenergies of topological edge states are equal to half the dissipation rate.
Our work reveals that the stationary entanglement can be a quantum signature of the topological phase in bosonic systems, and inversely the topological quadratic systems can be powerful platforms to generate robust entanglement.
\end{abstract}
\maketitle

\section{Introduction}

Quantum entanglement, a key feature of quantum effects, plays an important role in quantum information \cite{RevModPhys.81.865} and quantum metrology \cite{RevModPhys.90.035005}. Quantum entanglement allows two distant systems to be correlated with each other, and the measurement results of one system can influence that of the other system, which is in stark contrast to classical physics \cite{PhysRev.47.777,PhysRevX.13.021031}. Nowadays, quantum entanglement has been considered as a major quantum resource to realize quantum computational advantages \cite{arute_quantum_2019,zhong_quantum_2020,wu_strong_2021}. Moreover, entanglement in atomic ensembles can reduce the quantum noise with enhanced measurement sensitivity \cite{PhysRevLett.102.100401,gross_nonlinear_2010,riedel_atom-chip-based_2010,luo_deterministic_2017,hosten_measurement_2016,zou_beating_2018,pedrozo-penafiel_entanglement_2020,liu_nonlinear_2022,wu_quantum_2023}.

In the field of condensed matter physics, long-range entanglement is a signature of the quantum topological phase, which is a property of many-body systems with topological order \cite{RevModPhys.89.041004}. On the contrary, the topology widely investigated in ultracold atoms \cite{price_four-dimensional_2015,price_synthetic_2017,taddia_topological_2017,sugawa_second_2018,chalopin_probing_2020,wang_realization_2021,wang_evidence_2021} and photonic systems \cite{haldane_model_1988,lu_topological_2014,yang_realization_2019,el_hassan_corner_2019,li_higher-order_2020,ao_topological_2020,xia_nonlinear_2021,lustig_photonic_2022} is indeed classical topology, which originates from the geometric properties of the single-particle wave nature.
This kind of topology is characterized by robust edge states or topological invariants and is conventionally believed to be uncorrelated with quantum properties \cite{RevModPhys.91.015006}.

Parallelly, in the field of quantum optics, bosonic quadratic systems, which possess Hamiltonians that are quadratic in terms of bosonic creation and annihilation operators \cite{colpa_diagonalization_1978}, are an important method to generate quantum entanglement \cite{vitali_optomechanical_2007,tian_robust_2013,PhysRevLett.110.253601}.
The quadratic interactions exist in various platforms, such as bosonic fields with parametrically driving \cite{mittal_topological_2018,PhysRevLett.128.153603,sohn_topological_2022}, interacting Bose-Einstein condensate \cite{RevModPhys.78.179,PhysRevLett.93.140406,PhysRevX.9.011047,PhysRevX.10.011030} and optomechanical systems \cite{RevModPhys.86.1391,PhysRevA.88.053850}.
%Unlike conventional linear topological systems, quadratic squeezing interaction is also a major method to generate entanglement \cite{vitali_optomechanical_2007,tian_robust_2013,PhysRevLett.110.253601}.
Recently, it is shown that an open quadratic chain exhibits non-Hermitian dynamics \cite{mcdonald_phase-dependent_2018,yokomizo_non-bloch_2021,del_pino_non-hermitian_2022,wang_amplification_2022} and novel topology \cite{PhysRevLett.127.245701,PhysRevLett.130.123602,PhysRevLett.130.203605}. However, the relation between quantum entanglement and topology remains unclear in this system.

\begin{figure}[b]
	\centering
	\includegraphics[width=0.48\textwidth]{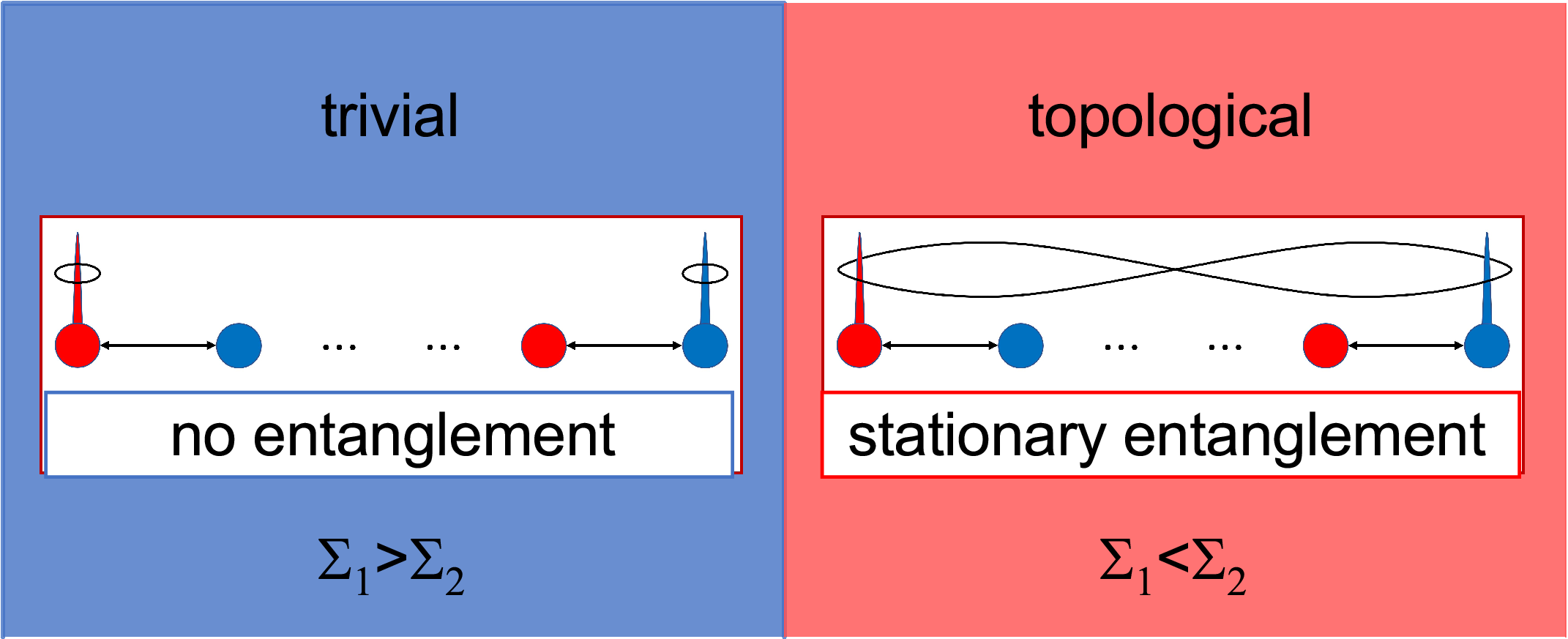}
	\caption{Stationary entanglement induced by bosonic topology. Phase diagram of the topology and the stationary entanglement with real parameters. The entanglement only emerges in the topological phase, while there is no entanglement in the trivial phase. The equations denote the parameter ranges of the trivial and the topological phases, see definition in Sec. \ref{sec:topo}.}
	\label{fig:a}
\end{figure}

Here we uncover the topology-induced entanglement effect in an open bosonic quadratic chain in the steady state. Such an entanglement only emerges between the two edge modes in the topological phase, while there is no entanglement in the trivial phase, as sketched in Fig. \ref{fig:a}.
%We show these stationary quantum behaviours follow a similar mechanism as a two-mode system.
The stationary entanglement is related to the coupling of the system to the environment quantum fluctuations and will be greatly suppressed if the system-environment couplings (denoted by the dissipation rate) do not match the intrasystem couplings (which determines the system eigenenergies).
As the absolute values of the eigenenergies of the topological edge states are much smaller than that of the bulk states, it offers the opportunity to match only the topological edge states with the system-environment coupling and thus selectly generate stationary entanglement between the topological edge states. Importantly, this kind of topological matching and related entanglements disappears in the trivial phase when there are no topological edge states.
To prove this idea, we approximately solve the Langevin equations by neglecting other eigenenergies except for the near-zero ones of the topological edge states, leading to analytical results, which perfectly match the numerical results obtained from the covariance approach based on the quantum master equations.
It is revealed that the emergence of the topological edge states can greatly enhance the squeezing correlations.
%especially between two edge modes. Specifically, we show the squeezing correlations reach a maximum when the absolute eigenenergies of the topological edge states are equal to half the dissipation rate, and thus it leads to obvious stationary entanglements that only exist in the topological phase.
%The stationary entanglements are modulated by the profiles of topological edge states and mainly exist between several edge modes.
Our work establishes a general relationship between classical topology and quantum entanglements, which sheds new light on the study of quantum topological photonics.
%which can be further applied to other kinds of lattices and also Floquet or higher-dimensional systems.

The rest of this work is organized as follows. In Sec. \ref{sec:topo}, we describe the system model of a bosonic quadratic chain and derive the topological phase transition through both the Bloch and non-Bloch band theory. In Sec. \ref{sec:qme}, we analyze the system through the covariance approach based on quantum master equations to obtain exact numerical results. In Sec. \ref{sec:qle}, we deduce approximate analytical results using the quantum Langevin equations. In Sec. \ref{sec:two}, we present the analytical results for the quantum behaviors in a two-mode system. In Sec. \ref{sec:trivial} and \ref{sec:topological}, we describe the quantum behaviors in the trivial and topological phases of the bosonic quadratic chain, respectively. In Sec. \ref{sec:tie}, we investigate the topology-induced entanglements between two edge modes. In Sec. \ref{sec:max}, we show how to understand the pattern of the logarithmic negativity and maximize the stationary entanglement through the analytical expressions. In Sec. \ref{sec:complex}, we discuss the stationary entanglements with complex-valued coupling strengths. In Sec. \ref{sec:exp}, we discuss the possible experimental realization. In Sec. \ref{sec:sum}, we conclude this work with some discussions. In the appendixes, we provide several parts of the detailed derivations, including the Bloch band theory (Appendix \ref{ap:b}), the non-Bloch band theory (Appendix \ref{ap:nb}), and the quantum Langevin equations (Appendix \ref{ap:qle}).

% (a) Bosonic quadratic chain with both staggered linear interactions ($t_{1}$, $t_{2}$) and squeezing interactions ($\Delta_{1}$, $\Delta_{2}$). The dotted box indicates the unit cell. A unit cell contains two bosonic modes with an equal dissipation rate $\kappa$. (b)
\section{Bosonic quadratic chain and topological phase transition}\label{sec:topo}

\begin{figure}[t]
	\centering
	\includegraphics[width=0.48\textwidth]{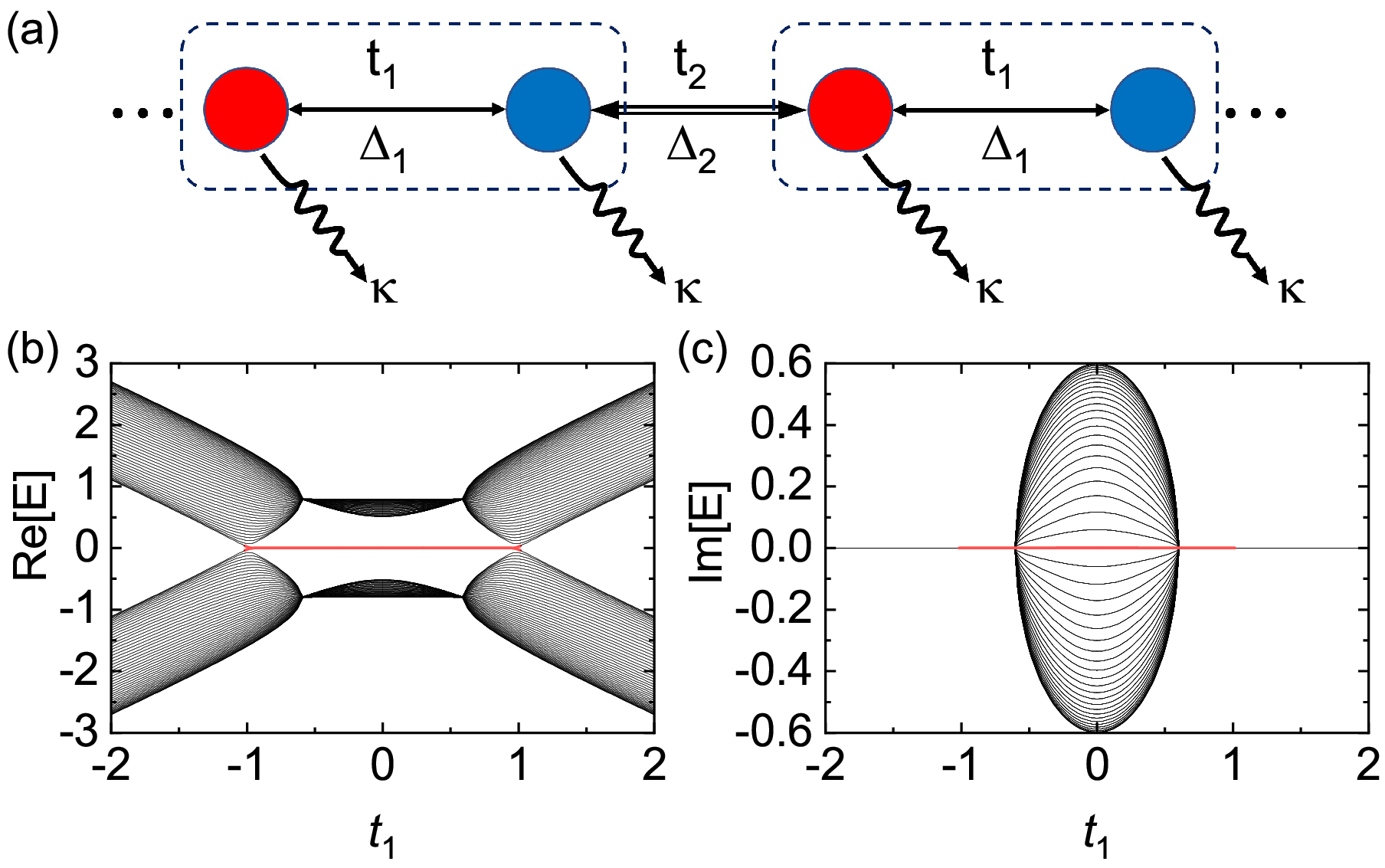}
	\caption{(a) Bosonic quadratic chain with both staggered linear interactions ($t_{1}$, $t_{2}$) and squeezing interactions ($\Delta_{1}$, $\Delta_{2}$). The dotted boxes indicate the unit cells, and each unit cell contains two bosonic modes. The wavy arrows denote the system-environment couplings described by the dissipation rate $\kappa$.
(b)-(c): The real (b) and imaginary (c) parts of energy spectra for an open chain with length $N=10$ (unit cell). The red lines indicate the topological edge modes at zero energy. Other parameters are $t_{\mathrm{2}}=0.8$, $\Delta_{\mathrm{\mathrm{1}}}=0.6$ and $\Delta_{\mathrm{2}}=0$.}
	\label{fig:2s}
\end{figure}

As depicted in Fig. \ref{fig:2s}(a), we consider a bosonic quadratic chain with both staggered linear interactions and squeezing interactions, which can be viewed as a generalization of the Su-Schrieffer-Heeger (SSH) model \cite{PhysRevLett.42.1698,PhysRevB.22.2099} by adding squeezing interactions. Moreover, we take the system-environment coupling into account by assuming that all the modes are coupled to a Markovian environment with a dissipation rate $\kappa$. The system Hamiltonian can be written as
\begin{equation}\label{eq:h1}
	\begin{split}
		H_{\mathrm{}}=&\sum_{j=1}^{N}(t_{\mathrm{}1}a_{2j-1}^{\dag} a_{2j}+\Delta_{\mathrm{}1} a_{2j-1}^{\dag} a_{2j}^{\dag}+\mathrm{H.c.})\\
		+&\sum_{j=1}^{N-1}(t_{\mathrm{}2} a_{2j+1}^{\dag} a_{2j}+\Delta_{\mathrm{}2} a_{2j+1}^{\dag} a_{2j}^{\dag}+\mathrm{H.c.}),
	\end{split}
\end{equation}
where $t_{\mathrm{}1}$ ($t_{\mathrm{}2}$) and $\Delta_{\mathrm{}1}$ ($\Delta_{\mathrm{}2}$) are the intracell (intercell) coupling strengths of linear and squeezing interactions, respectively, $N$ is the number of unit cells, and $a_{j}$ is the annihilation operator of the $j$th mode. The Bloch Hamiltonian of the system can be written as
\begin{equation}
	\mathcal{H}_{\mathrm{}}(k)=(t_{\mathrm{}1}+t_{\mathrm{}2}e^{ik})a_{k}^{\dag}a_{k}^{\prime}+(\Delta_{\mathrm{}1}+\Delta_{\mathrm{\mathrm{}2}}e^{ik})a_{k}^{\dag}a_{-k}^{\prime\dag}+\mathrm{H.c.},
\end{equation}
or in the matrix form: $\mathcal{H}(k)=\frac{1}{2}C_{k}^{\dag}\mathcal{H}_{\mathrm{M}}(k)C_{k}$, where $C_{k}^{\dag}=(a_{k}^{\dag},a_{k}^{\prime\dag},a_{-k},a_{-k}^{\prime})$ [see Appendix \ref{ap:b} for details]. The system satisfies the chiral symmetry, i.e., $\Gamma \mathcal{H}_{\mathrm{M}}\Gamma = -\mathcal{H}_{\mathrm{M}}$ for $\Gamma=\sigma_{3}\otimes\sigma_{0}$, where $\sigma_{3}$ is the third Pauli matrix and $\sigma_{0}$ is two-dimensional identity matrix.

When the quadratic squeezing terms are nonzero, the excitation modes are no longer unitary transformations of initial bosonic modes. Instead, the excitation modes are Bogoliubov modes that are determined by the eigenvalue equation of $\tau_{z}\mathcal{H}_{\mathrm{M}}$, where $\tau_{z}=\mathrm{Diagonal}(\mathbbm{1},-\mathbbm{1})$ and $\mathbbm{1}$ is an identity matrix with half the dimension of $\mathcal{H}_{\mathrm{M}}$ \cite{yokomizo_non-bloch_2021}.

Without loss of generality, we assume $t_{\mathrm{}2}=q_{\mathrm{t}}t_{\mathrm{1}}e^{i\phi_{\mathrm{t}}}$, $\Delta_{\mathrm{}2}=q_{\mathrm{\Delta}}\Delta_{\mathrm{}1}e^{i\phi_{\Delta}}$ and $t_{\mathrm{1}}$, $\Delta_{1}$, $q_{\mathrm{t},\Delta}$ are all real. Then the eigenvalues of $\tau_{z} \mathcal{H}_{\mathrm{M}}(k)$ can be obtained as
\begin{equation}\label{eq:lambda}
	\xi^{2}=\Sigma_{1}+\Sigma_{2}+2\Sigma_{3}\cos k\pm 2|\sin k|\sqrt{\Sigma_{1}\Sigma_{2}-\Sigma_{3}^{2}},
\end{equation}
where $\Sigma_{1}=t_{\mathrm{}1}^{2}-\Delta_{\mathrm{}1}^{2}$, $\Sigma_{2}=q_{\mathrm{t}}^{2}t_{\mathrm{}1}^{2}-q_{\Delta}^{2}\Delta_{\mathrm{}1}^{2}$ and $\Sigma_{3}=q_{\mathrm{t}}\cos \phi_{\mathrm{t}}t_{\mathrm{}1}^{2}-q_{\Delta}\cos \phi_{\Delta}\Delta_{\mathrm{}1}^{2}$. For $\Sigma_{1}\Sigma_{2}-\Sigma_{3}^{2}\ge0$, the Bloch spectrum is real, indicating that there is no non-Hermitian skin effect. We can directly obtain the energy spectrum as
\begin{equation}\label{eq:band}
    (\sqrt{\Sigma_{1}}-\sqrt{\Sigma_{2}})^{2}<\xi^{2}<(\sqrt{\Sigma_{1}}+\sqrt{\Sigma_{2}})^{2},
\end{equation}
where we assume $\Sigma_{1,2}$ are both positive. When $\Sigma_{1}<0$ or $\Sigma_{2}<0$, the eigenvalues of the long open chain are imaginary, and the system becomes unstable. Consequently, we are only interested in the stable region for $\Sigma_{1,2}>0$. According to the bulk-edge correspondence, the gap-closing points $\Sigma_{1}=\Sigma_{2}$ ($|t_{1}|^{2}-|\Delta_{1}|^{2}=|t_{2}|^{2}-|\Delta_{2}|^{2}$) also denote the gap-closing points of an open chain and are where the topological phase transition takes place.

However, when $\Sigma_{1}\Sigma_{2}-\Sigma_{3}^{2}<0$, the Bloch spectrum forms a loop in the complex energy plane with the emergence of the non-Hermitian skin effect. In this case, the Bloch bulk-edge correspondence fails but can be rebuilt with the non-Bloch theory. The non-Bloch matrix $\mathcal{H}_{\mathrm{M}}(\beta)$ can be obtained from $\mathcal{H}_{\mathrm{M}}(k)$ by the replacements $e^{ik}\to\beta$, $e^{-ik}\to\beta^{-1}$ \cite{yokomizo_non-bloch_2021}, which is
\begin{equation}
	\begin{split}
		\mathcal{H}_{\mathrm{}}(\beta)=&(t_{\mathrm{}1}+t_{\mathrm{}2}\beta)a_{k}^{\dag}a_{k}^{\prime}+(\Delta_{\mathrm{}1}+\Delta_{\mathrm{\mathrm{}2}}\beta)a_{k}^{\dag}a_{-k}^{\prime\dag}\\
		+&(t_{\mathrm{}1}^{*}+t_{\mathrm{}2}^{*}\beta^{-1})a_{k}^{}a_{k}^{\prime\dag}+(\Delta_{\mathrm{}1}^{*}+\Delta_{\mathrm{\mathrm{}2}}^{*}\beta^{-1})a_{k}^{}a_{-k}^{\prime}.
	\end{split}
\end{equation}
The eigenvalues of $\tau_{z} \mathcal{H}_{\mathrm{M}}(\beta)$ become
\begin{equation}\label{eq:lambda2}
	\begin{split}
		\xi^{2}=&\Sigma_{1}+\Sigma_{2}+\Sigma_{3}(\beta+\beta^{-1})\\
		\pm& \sqrt{-(\beta-\beta^{-1})^{2}}\sqrt{\Sigma_{1}\Sigma_{2}-\Sigma_{3}^{2}}.
	\end{split}
\end{equation}
We can also obtain the generalized momentum as
\begin{equation}\label{eq:beta}
	\beta=\frac{1}{2}\frac{\lambda^{2}-\Sigma_{1}-\Sigma_{2}\pm\sqrt{(\xi^{2}-\Sigma_{1}-\Sigma_{2})^{2}-4\Sigma_{1}\Sigma_{2}}}{\Sigma_{3}\pm \sqrt{\Sigma_{3}^{2}-\Sigma_{1}\Sigma_{2}}}.
\end{equation}
There are two ``$\pm$'' and four $\beta$. The four $\beta$ are two pairs according to the $\pm$ in the denominator. We note the denominator is real as the term under the root sign is positive. The existence of the generalized Brillouin zone requires the absolute values of two $\beta$ in each pair equal to each other. It means the term under the root sign in the numerator is negative, i.e., [see Appendix \ref{ap:nb} for details]
\begin{equation}
	(\xi^{2}-\Sigma_{1}-\Sigma_{2})^{2}-4\Sigma_{1}\Sigma_{2}<0.
\end{equation}

Interestingly, the non-Bloch Hamiltonian also gives the same energy spectrum as Eq. (\ref{eq:band}). It means the topological phase transition also takes place at $\Sigma_{1}=\Sigma_{2}$ in the case with the non-Hermitian skin effect. The open chain is in the topological phase for $\Sigma_{1}<\Sigma_{2}$ and in the trivial phase for $\Sigma_{1}>\Sigma_{2}$. We note the same energy spectrum is a coincidence. The conventional bulk-boundary correspondence still fails as the Bloch spectrum can not provide the open-boundary spectrum.
%  \cite{note_nonbloch}.

Figure \ref{fig:2s}(b) and \ref{fig:2s}(c) are typical energy spectra of the quadratic system with the emergence of the non-Hermitian skin effect. The energy spectra is all real when $|t_{1}/\Delta_{1}|>1$, although the Bloch spectrum can be complex. We observe a topological phase transition at $t_{1}=1=\sqrt{|t_{2}|^{2}-|\Delta_{2}|^{2}+|\Delta_{1}|^{2}}$, as predicted by the non-Bloch theory.

\section{Covariance approach based on quantum master equations}\label{sec:qme}

We focus on the stationary quantum behaviors of the system, which can be obtained by calculating the time evolution of the system and taking the long-time limits or directly calculating the time-independent equilibrium solutions. In this section, we use the covariance approach based on quantum master equations to obtain exact numerical results.

The quantum master equation is given by $\dot{\rho}=i[\rho,H]+\kappa (1+n_{\mathrm{th}})\sum_{j=1}^{2N}\mathcal{D}(a_{j})\rho+\kappa n_{\mathrm{th}}\sum_{j=1}^{2N}\mathcal{D}(a_{j}^{\dag})\rho$, where $\mathcal{D}(\hat{o})\rho=\hat{o}\rho\hat{o}^{\dag}-(\hat{o}^{\dag}\hat{o}\rho+\rho\hat{o}^{\dag}\hat{o})/2$ is the Liouvillian for operator $\hat{o}$, and $n_{\mathrm{th}}$ is the environment photon number. 
This equation gives all the information of the density matrix, but the Hilbert space of bosonic systems is infinity, so the truncation process is required, and it still consumes too much computational resources.

To capture the most important features of quantum correlations, we only need to consider the covariances (second-order moments) $\langle\hat{o}\hat{o}^{\prime}\rangle$, where $\hat{o},\hat{o}^{\prime}\in\{a_{j},a_{j}^{\dag},j=1,2,\cdots, N\}$ are either an annihilation or creation operator. 
By using this covariance approach we can obtain exact numerical results without truncation process  \cite{liu_dynamic_2013}.
The evolution equations of the second-order moments can be obtained from the quantum master equations $d\langle\hat{o}\hat{o}^{\prime}\rangle/dt=\mathrm{Tr}(\dot{\rho}\hat{o}\hat{o}^{\prime})$, which allow us to numerically analyze both the dynamic and stationary behaviors of the system.
To obtain the stationary mean values of the second-order moments, we can let the time derivations equal to zero. Specifically, we are interested in the entanglement between two edge modes $a_{1}$ and $a_{2N}$. Then the degree of the two-mode entanglement can be quantified by the logarithmic negativity $E_{N}$, which is a function of the covariance matrix of the two modes \cite{PhysRevA.65.032314,PhysRevLett.110.253601}.

\section{Analytical results through quantum Langevin equations}\label{sec:qle}

Although the exact numerical results can be obtained using the approach in the previous section, the underlying physical mechanism is hard to analyze. In this section we calculate the quantum Langevin equations which provide an approximate route to capture the physical mechanism analytically.

From the original system Hamiltonian in Eq. (\ref{eq:h1}), we can find that the Langevin equations of a quadratic system include both the annihilation and creation operators, which is difficult to be solved analytically. To overcome this problem, we employ a squeezing transformation to transform the quadratic Hamiltonian into a Hamiltonian without the quadratic interactions \cite{mcdonald_phase-dependent_2018}, and the squeezing property now is transform to the
noise operators.
For simplicity in calculation, we first rewrite the system Hamiltonian in the quadrature representation [$a_{j}=(x_{j}+ip_{j})/\sqrt{2}$], which is
\begin{equation}
	\begin{split}
		H=&\sum_{j=1}^{N}\left[(t_{1}+\Delta_{1})x_{2j-1}x_{2j}+(t_{1}-\Delta_{1})p_{2j-1}p_{2j}\right]\\
		+&\sum_{j=1}^{N-1}\left[(t_{2}+\Delta_{2})x_{2j+1}x_{2j}+(t_{2}-\Delta_{2})p_{2j+1}p_{2j}\right].
	\end{split}
\end{equation}
We employ the squeezing transformation $x_{j}=e^{-r_{j}}\tilde{x}_{j}$ and $p_{j}=e^{r_{j}}\tilde{p}_{j}$, then the quadratic Hamiltonian becomes the Hamiltonian of a simple SSH chain
\begin{equation}\label{eq:sshp}
	\begin{split}
		\tilde{H}=&\sum_{j=1}^{N}t_{1}^{\prime}(\tilde{x}_{2j-1}\tilde{x}_{2j}+\tilde{p}_{2j-1}\tilde{p}_{2j})\\
		&+\sum_{j=1}^{N-1}t_{2}^{\prime}(\tilde{x}_{2j}\tilde{x}_{2j+1}+\tilde{p}_{2j}\tilde{p}_{2j+1})\\
		=&\sum_{j=1}^{N}t_{1}^{\prime}\tilde{a}_{2j-1}^{\dag}\tilde{a}_{2j}+\sum_{j=1}^{N-1}t_{2}^{\prime}\tilde{a}_{2j+1}^{\dag}\tilde{a}_{2j}+\mathrm{H.c.},
	\end{split}
\end{equation}
where $t_{j}^{\prime}=\sqrt{t_{j}^{2}-\Delta_{j}^{2}}$ for $j=1,2$. The site-dependent squeezing parameters in the squeezing transformation are given by
% $r_{2j-1}=(j-1)(r_{\mathrm{b}}-r_{\mathrm{a}})+r_{0}$ and $r_{2j}=-j(r_{\mathrm{b}}-r_{\mathrm{a}})+r_{\mathrm{b}}-r_{0}$,
\begin{gather}
	r_{2j-1}=(j-1)(r_{\mathrm{b}}-r_{\mathrm{a}})+r_{0},\\
	r_{2j}=-j(r_{\mathrm{b}}-r_{\mathrm{a}})+r_{\mathrm{b}}-r_{0},
\end{gather}
where $e^{2r_{\mathrm{a}}}={(t_{1}+\Delta_{1})}/{(t_{1}-\Delta_{1})}$ and $e^{2r_{\mathrm{b}}}={(t_{2}+\Delta_{2})}/{(t_{2}-\Delta_{2})}$. $r_{0}$ is a constant that can be arbitrarily chosen in the squeezing transformation. Here we determine it through the mirror symmetry, i.e., $r_{1}=r_{2N}$, which can simplify the calculation of the Langevin equations.

Then we can obtain the Langevin equations of the new operators as
\begin{equation}\label{eq:langevin1}
	\dot{\tilde{a}}_{2j-1}=-\frac{\kappa}{2}\tilde{a}_{2j-1}-it_{1}^{\prime}\tilde{a}_{2j}-it_{2}^{\prime}\tilde{a}_{2j-2}-\sqrt{\kappa}\tilde{a}_{\mathrm{in},2j-1},
\end{equation}
\begin{equation}\label{eq:langevin2}
	\dot{\tilde{a}}_{2j}=-\frac{\kappa}{2}\tilde{a}_{2j}-it_{1}^{\prime}\tilde{a}_{2j-1}-it_{2}^{\prime}\tilde{a}_{2j+1}-\sqrt{\kappa}\tilde{a}_{\mathrm{in},2j},
\end{equation}
where $\tilde{a}_{\mathrm{in},j}$ are the noise operators. Due to the squeezing transformation, these noise operators denote couplings to a squeezed environment. The above Langevin equations can be rewritten in the matrix form as
\begin{equation}
	\dot{\tilde{\bm{\mathrm{A}}}}=(-\frac{\kappa}{2}\mathbbm{1}-iS)\tilde{\bm{\mathrm{A}}}-\sqrt{\kappa}\tilde{\bm{\mathrm{A}}}_{\mathrm{in}},
\end{equation}
where $\tilde{\bm{\mathrm{A}}}=(\tilde{a}_{1},\cdots)^{\mathrm{T}}$, $\tilde{\bm{\mathrm{A}}}_{\mathrm{in}}=(\tilde{a}_{1,\mathrm{in}},\cdots)^{\mathrm{T}}$, $\mathbbm{1}$ is the identity matrix and $S$ is the coupling matrix [see Appendix \ref{ap:qle} for details]. Importantly, after the squeezing transformation, the coupling matrix $S$ is Hermitian and can be diagonalized as $S=PJP^{-1}$. $P=(\bm{\alpha}_{1},\bm{\alpha}_{2},\cdots)$ and the column vectors $\alpha_{j}$ are the eigenvectors of $S$. The diagonal elements of the diagonal matrix $J=\mathrm{Diag}({\lambda_{1},\lambda_{2},\cdots})$ is the corresponding eigenvalues. Then we can obtain the stationary solutions as
\begin{equation}
	\tilde{\bm{\mathrm{A}}}_{\mathrm{s}}=\sqrt{\kappa}\sum_{j=1}^{2N}\lim_{t\to\infty}\int_{0}^{t}e^{(-\frac{\kappa}{2}-i\lambda_{j})(t-t^{\prime})}[\bm{\alpha}_{j}\cdot \tilde{\bm{\mathrm{A}}}_{\mathrm{in}}(t^{\prime})]\bm{\alpha}_{j}dt^{\prime},
\end{equation}
or
\begin{equation}
	\tilde{a}_{m,\mathrm{s}}=\sqrt{\kappa}\sum_{j,k}\lim_{t\to\infty}\int_{0}^{t}e^{(-\frac{\kappa}{2}-i\lambda_{j})(t-t^{\prime})}\alpha_{j,k}\alpha_{j,m}\tilde{a}_{\mathrm{in},k}(t^{\prime})dt^{\prime}.
\end{equation}

Following the stationary solutions, the mean values of the second-order moments can be obtained as %($n_{\mathrm{th}}=0$)
\begin{equation}\label{eq:aa1}
	\langle \tilde{a}_{m}^{\dag}\tilde{a}_{m^{\prime}}\rangle_{\mathrm{s}}=\sum_{j,k,j^{\prime}}\frac{\kappa\alpha_{j,k}\alpha_{j,m}\alpha_{j^{\prime},k}\alpha_{j^{\prime},m^{\prime}}}{\kappa+i(-\lambda_{j}^{*}+\lambda_{j^{\prime}})}\frac{e^{2r_{k}}+e^{-2r_{k}}-2}{4},
\end{equation}
\begin{equation}\label{eq:aa2}
	\langle \tilde{a}_{m}\tilde{a}_{m^{\prime}}\rangle_{\mathrm{s}}=\sum_{j,k,j^{\prime}}\frac{\kappa\alpha_{j,k}\alpha_{j,m}\alpha_{j^{\prime},k}\alpha_{j^{\prime},m^{\prime}}}{\kappa+i(\lambda_{j}+\lambda_{j^{\prime}})}\frac{e^{2r_{k}}-e^{-2r_{k}}}{4}.
\end{equation}
Here and below the summation range is from $1$ to $2N$ if there is no additional description. For simplicity, we have assumed the environment photon number $n_{\mathrm{th}}=0$, and the full expressions can be found in Appendix \ref{ap:qle}.

\section{Hint from the two-mode system}\label{sec:two} %Quantum behaviors of a two-mode system

In this section we analyze the quantum behaviors of a two-mode system, which is a special case of $N=1$ and can be solved analytically without approximation, so that we can obtain some hints on the quantum entanglement generation. 
Following the derivation in Sec. \ref{sec:qle}, for two-mode system the squeezing parameters are $r_{1}=r_{2}=r_{0}=r_{\mathrm{a}}/2$. The eigenvalues and the eigenvectors are $\lambda_{1}=-\lambda_{2}=t_{1}^{\prime}$ and $\bm{\alpha}_{1}=(1/\sqrt{2},1/\sqrt{2})^{\mathrm{T}}$, $\bm{\alpha}_{2}=(1/\sqrt{2},-1/\sqrt{2})^{\mathrm{T}}$. So the mean values of the second-order moments are
\begin{equation}
	\langle \tilde{a}_{1}^{\dag}\tilde{a}_{1}\rangle_{\mathrm{s}}=\langle \tilde{a}_{2}^{\dag}\tilde{a}_{2}\rangle_{\mathrm{s}}=\frac{e^{r_{\mathrm{a}}}+e^{-r_{\mathrm{a}}}-2}{4},
\end{equation}
\begin{equation}
	\langle \tilde{a}_{1}^{\dag}\tilde{a}_{2}\rangle_{\mathrm{s}}=0,
\end{equation}
\begin{equation}
	\langle \tilde{a}_{1}^{2}\rangle_{\mathrm{s}}=\langle \tilde{a}_{2}^{2}\rangle_{\mathrm{s}}=\frac{\kappa^{2}}{\kappa^{2}+4t_{1}^{\prime2}}\frac{e^{r_{\mathrm{a}}}-e^{-r_{\mathrm{a}}}}{4},
\end{equation}
\begin{equation}\label{eq:twos}
	\langle \tilde{a}_{1}\tilde{a}_{2}\rangle_{\mathrm{s}}=-\frac{2i\kappa t_{1}^{\prime}}{\kappa^{2}+4t_{1}^{\prime2}}\frac{e^{r_{\mathrm{a}}}-e^{-r_{\mathrm{a}}}}{4}.
\end{equation}

From the squeezing correlation term Eq. (\ref{eq:twos}) we can find that the quantum correlation depends on two factors. The first factor is the squeezing parameter $r_{\mathrm{a}}$. The squeezing correlation term becomes zero when $r_{\mathrm{a}}=0$, i.e. $\Delta_{1}= 0$, which reveals that the existing of squeezing interaction is necessary for the emergence of stationary correlations. The second factor is the ratio between the dissipation rate $\kappa$ and the effective coupling strength $t_{1}^{\prime}$.
According to the fluctuation-dissipation theorem, the dissipation of a system is always connected to the noise fluctuation from the environment,
%The dissipation rate $\kappa$ represents two kinds of processes: the loss of amplitude or intensity and the coupling to an environment with quantum fluctuation. 
so both processes correspond to the same parameter $\kappa$ denoting system-environment coupling, which appears both at the denominator and numerator in Eq. (\ref{eq:twos}).
When $\kappa \ll t_{1}^{\prime}$, the squeezing correlation will be suppressed because the the coupling to the environment fluctuation is weak.
On the other hand, when $\kappa \gg t_{1}^{\prime}$, the strong dissipation will also suppress the squeezing correlation. Therefore, the optimal squeezing correlation is obtained for a moderate $\kappa/t_{1}^{\prime}$, which means that the system-environment coupling should match the intrasystem coupling.
From Eq. (\ref{eq:twos}) we can find that the optimal condition is $\kappa=2t_{1}^{\prime}$.

The above analysis for the two-mode system provides the physical insights for a bosonic chain with more modes. In this case the energy levels become energy bands, thus it is natural to consider the effect of eigenenergies. We can infer that the eigenenergies should match the system-environment coupling to obtain optimal squeezing correlation.
In the topological phase, there exists topological edge states whose eigenenergies are near zero and separated from the bulk energy bands, thus it offers the opportunity to generate long-range entanglement between two edge modes when the corresponding eigenenergies match the system-environment coupling with near zero $\kappa$, while at the same time the squeezing correlations between bulk states are suppressed.

\section{Quantum behaviors in the trivial phase}\label{sec:trivial}

In this section, we will prove that the squeezing correlations in a bosonic quadratic chain with trivial phase are greatly suppressed for small $\kappa$ (compared to the coupling strength), and there are no stationary entanglements in this case. In the trivial phase for $\Sigma_{1}>\Sigma_{2}$, the lattice spectrum opens a trivial gap, and all the eigenvalues have finite absolute values which we assume are much larger than the dissipation rate $|\lambda_{j}|\gg\kappa$. So in the summations Eq. (\ref{eq:aa1})-(\ref{eq:aa2}), those terms with eigenvalues that cancel out with each other are much larger than other terms. Appropriately, we only consider these large terms, and the summations become
\begin{equation}\label{eq:gg1}
	\langle \tilde{a}_{m}^{\dag}\tilde{a}_{m^{\prime}}\rangle_{\mathrm{s}}\approx\sum_{k,j}\alpha_{j,k}^{2}\alpha_{j,m}\alpha_{j,m^{\prime}}\frac{e^{2r_{k}}+e^{-2r_{k}}-2}{4},
\end{equation}
\begin{equation}\label{eq:ff1}
	\langle \tilde{a}_{m}\tilde{a}_{m^{\prime}}\rangle_{\mathrm{s}}\approx\sum_{k,j}\alpha_{j,k}\alpha_{j,m}\alpha_{j,k}^{\prime}\alpha_{j,m^{\prime}}^{\prime}\frac{e^{2r_{k}}-e^{-2r_{k}}}{4},
\end{equation}
where $\alpha_{j,k}^{\prime}$ denotes the eigenvector with an opposite eigenvalue of $\alpha_{j,k}$. As this system preserves chiral symmetry, the pair of eigenvectors with opposite eigenvalues satisfy $\alpha_{j,2k-1}=\alpha_{j,2k-1}^{\prime}$, $\alpha_{j,2k}=-\alpha_{j,2k}^{\prime}$. Moreover, the system preserves mirror symmetry $|\alpha_{j,k}|=|\alpha_{j,2N+1-k}|$. In the meanwhile, the squeezing parameters satisfy $r_{k}=r_{2N+1-k}$. Consequently, every term in the summation Eq. (\ref{eq:ff1}) is zero. For a similar reason, the summations in Eq. (\ref{eq:aa1}) are equal to zero when $m$ and $m^{\prime}$ are not both odd or even, as every pair of terms with opposite eigenvalues cancel out ($\alpha_{j,k}^{2}\alpha_{j,m}\alpha_{j,m^{\prime}}+\alpha_{j,k}^{\prime2}\alpha_{j,m}^{\prime}\alpha_{j,m^{\prime}}^{\prime}=0$). It means the only non-zero terms are those like $\langle \tilde{a}_{2m}^{\dag}\tilde{a}_{2m^{\prime}}\rangle_{\mathrm{s}}$ and $\langle \tilde{a}_{2m+1}^{\dag}\tilde{a}_{2m^{\prime}+1}\rangle_{\mathrm{s}}$.

These properties mean the total lattice can be divided into two sublattices: the odd modes and the even modes. The modes between two sublattices have no quantum correlation. Moreover, as the squeezing correlation terms ($m\neq m^{\prime}$) or single-mode squeezing terms ($m = m^{\prime}$) in Eq. (\ref{eq:ff1}) are always zero, there is no quantum squeezing effect in the squeezing representation. In other words, the squeezing parameters $r_{j}$ in the squeezing transformation are exactly the squeezing coefficient of every mode in the steady state.

\section{Quantum behaviors in the topological phase}\label{sec:topological}
%  ($|t_{1}|^{2}-|\Delta_{1}|^{2}<|t_{2}|^{2}-|\Delta_{2}|^{2}$)

In the topological phase for $\Sigma_{1}<\Sigma_{2}$, the energy spectrum is different from that in the trivial phase, with the emergence of topological edge states. The absolute energy of two edge states $|\lambda_{1,2}|$ are much smaller than the absolute values of other eigenvalues. Consequently, the contribution of the topological edge states must be considered in the summations. For simplicity, we note that in this work the concept of ``state" denotes the eigenstates of the chain, while the concept of ``mode" denotes the original physical modes in the chain. The summations of Eq. (\ref{eq:aa1})-(\ref{eq:aa2}) become
\begin{equation}\label{eq:aaf1}
	\begin{split}
		\langle \tilde{a}_{m}^{\dag}&\tilde{a}_{m^{\prime}}\rangle_{\mathrm{s}}\approx\sum_{k,j}\alpha_{j,k}^{2}\alpha_{j,m}\alpha_{j,m^{\prime}}\frac{e^{2r_{k}}+e^{-2r_{k}}-2}{4}\\
		+&\sum_{k \atop j=1,2}\frac{\kappa}{\kappa+2i\lambda_{j}}\alpha_{j,k}\alpha_{j,m}\alpha_{j,k}^{\prime}\alpha_{j,m^{\prime}}^{\prime}\frac{e^{2r_{k}}+e^{-2r_{k}}-2}{4},
	\end{split}
\end{equation}
\begin{equation}\label{eq:aag1}
	\begin{split}
		\langle \tilde{a}_{m}\tilde{a}_{m^{\prime}}\rangle_{\mathrm{s}}&\approx\sum_{k,j}\alpha_{j,k}\alpha_{j,m}\alpha_{j,k}^{\prime}\alpha_{j,m^{\prime}}^{\prime}\frac{e^{2r_{k}}-e^{-2r_{k}}}{4}\\
		+&\sum_{k \atop j=1,2}\frac{\kappa}{\kappa+2i\lambda_{j}}\alpha_{j,k}^{2}\alpha_{j,m}\alpha_{j,m^{\prime}}\frac{e^{2r_{k}}-e^{-2r_{k}}}{4}.
	\end{split}
\end{equation}
% , as the terms for $k$ and $2N+1-k$ are opposite
As proved in the previous section, the first line in Eq. (\ref{eq:aag1}) is zero. Similarly, the second line in Eq. (\ref{eq:aaf1}) is also zero. Then the summation Eq. (\ref{eq:aaf1}) returns to Eq. (\ref{eq:gg1}), but the squeezing correlation terms or single-mode squeezing terms Eq. (\ref{eq:aag1}) keep non-zero as
\begin{equation}\label{eq:aag2}
	\langle \tilde{a}_{m}\tilde{a}_{m^{\prime}}\rangle_{\mathrm{s}}\approx\sum_{k \atop j=1,2}\frac{\kappa}{\kappa+2i\lambda_{j}}\alpha_{j,k}^{2}\alpha_{j,m}\alpha_{j,m^{\prime}}\frac{e^{2r_{k}}-e^{-2r_{k}}}{4}.
\end{equation}

We point out that Eq. (\ref{eq:aag2}) summarizes the key analytical results of this work. The non-zero squeezing correlation terms in Eq. (\ref{eq:aag2}) lead to the emergence of two-mode entanglement in the steady state, and the single-mode squeezing terms in Eq. (\ref{eq:aag2}) lead to a modulation of squeezing degree and squeezing phase in every mode. Importantly, unlike the terms in Eq. (\ref{eq:gg1}) which are non-zero only when $m$ and $m^{\prime}$ are both odd or even, the terms in Eq. (\ref{eq:aag2}) are always non-zero irrespective of $m$ and $m^{\prime}$. It denotes that there are also quantum correlations between modes in two sublattices, which do not exist in the trivial phase. Moreover, as shown in the derivation, the non-zero terms in Eq. (\ref{eq:aag2}) originate from the near-zero energies of two topological edge modes. So the quantum effects such as the quantum entanglements can be viewed as the quantum signatures of the topological edge modes.

We assume the eigenvalues of two edge states are $\lambda_{1}=\delta$ and $\lambda_{2}=-\delta$. The distributions of two edge states can be approximately given by $\alpha_{1,2j-1}=\alpha_{2,2j-1}\approx le^{-(j-1)\varepsilon}$ and $\alpha_{1,2j}=-\alpha_{2,2j}\approx le^{(j-N)\varepsilon}$, where $\varepsilon\approx \ln t_{2}^{\prime}-\ln t_{1}^{\prime}$ is the topological localization coefficient, and $l=1/\sqrt{2(1-e^{-2(N-1)\varepsilon})/(1-e^{-2\varepsilon})}$ is the normalization coefficient \cite{asboth_short_2016}. Then Eq. (\ref{eq:aag2}) can be reduced to
\begin{equation}\label{eq:easy21}
	\langle \tilde{a}_{2m}\tilde{a}_{2m^{\prime}}\rangle_{\mathrm{s}}\approx\frac{\kappa^{2} l^{4}e^{(m+m^{\prime}-2N)\varepsilon}}{\kappa^{2}+4\delta^{2}}\left[e^{-2r_{0}}L_{1}-e^{2r_{0}}L_{2}\right],
\end{equation}
\begin{equation}\label{eq:easy22}
	\langle \tilde{a}_{2m+1}\tilde{a}_{2m^{\prime}+1}\rangle_{\mathrm{s}}\approx\frac{\kappa^{2} l^{4}e^{-(m+m^{\prime}-2)\varepsilon}}{\kappa^{2}+4\delta^{2}}\left[e^{-2r_{0}}L_{1}-e^{2r_{0}}L_{2}\right],
\end{equation}
\begin{equation}\label{eq:easy31}
	\begin{split}
		\langle \tilde{a}_{2m}&\tilde{a}_{2m^{\prime}+1}\rangle_{\mathrm{s}}\approx\\
		-&\frac{2i\delta\kappa l^{4}e^{(m-m^{\prime}-N+1)\varepsilon}}{\kappa^{2}+4\delta^{2}}\left[e^{-2r_{0}}L_{1}-e^{2r_{0}}L_{2}\right],
	\end{split}
\end{equation}
where $L_{1}=[1-e^{-2(N-1)(\varepsilon+r_{\mathrm{b}}-r_{\mathrm{a}})}]/[1-e^{-2(\varepsilon+r_{\mathrm{b}}-r_{\mathrm{a}})}]$ and $L_{2}=[1-e^{-2(N-1)(\varepsilon-r_{\mathrm{b}}+r_{\mathrm{a}})}]/[1-e^{-2(\varepsilon-r_{\mathrm{b}}+r_{\mathrm{a}})}]$.

\begin{figure}[t]
	\centering
	\includegraphics[width=0.45\textwidth]{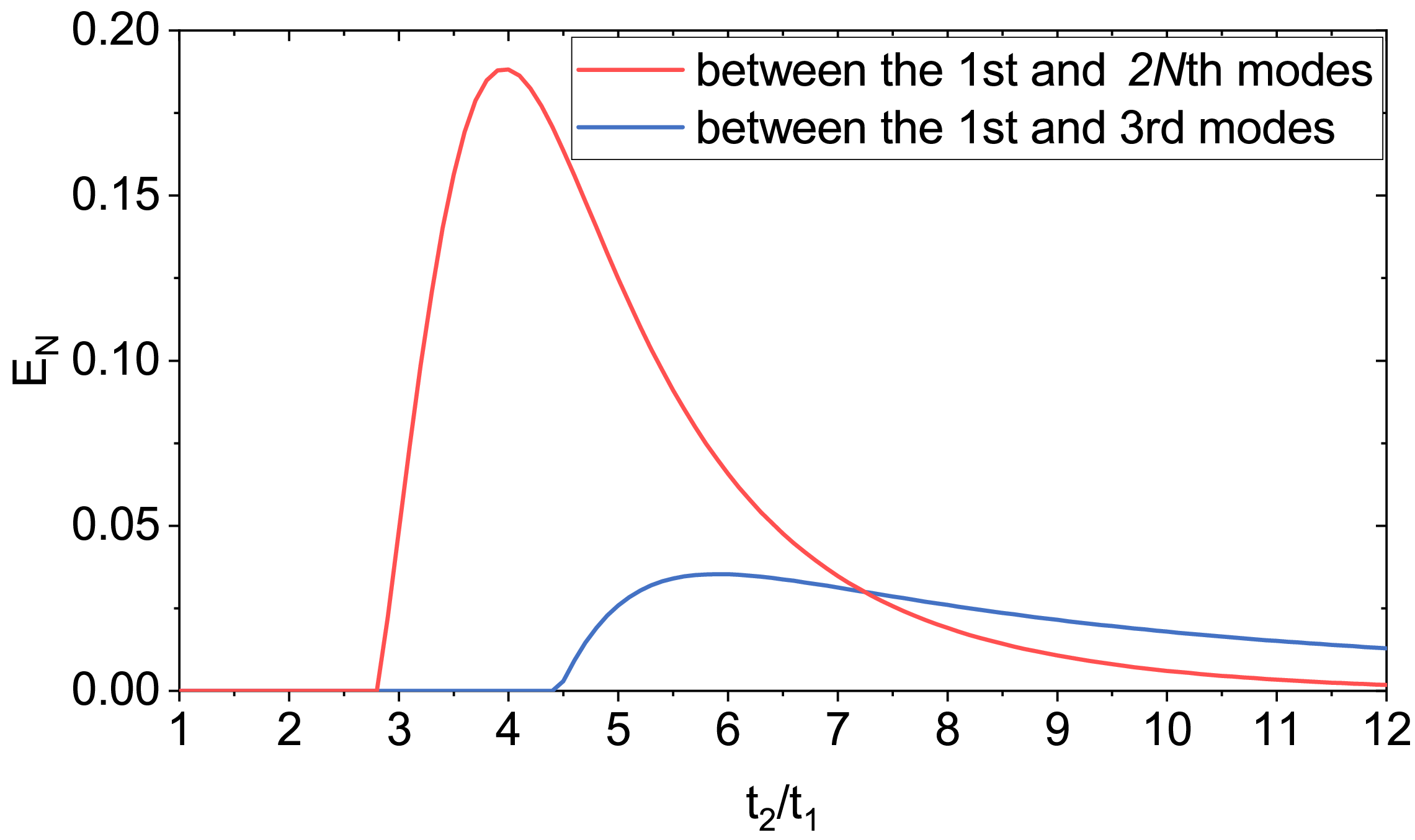}
	\caption{Stationary entanglement (quantified by $E_{N}$) between two edge modes (red) and between the first modes and the third mode (blue) versus the ratio of linear coupling strengths $t_{2}/t_{1}$. Other parameters are $\kappa=0.01$, $N=5$, $\mu=0$, and $\Delta_{1}/t_{1}=\Delta_{2}/t_{2}=0.6$.}
	\label{fig:2}
\end{figure}

All the above squeezing terms are modulated by the exponential distribution of the topological edge state, and these terms decrease quickly when considering modes far away from the edges. In Fig. \ref{fig:2}, we plot the stationary entanglement (quantified by $E_{N}$) between two edge modes (red) and between the first mode and the third mode (blue) versus the ratio of linear coupling strengths $t_{2}/t_{1}$. The maximal stationary entanglement in the latter case is much smaller than in the former case. Moreover, we find that there is no stationary entanglement between other pairs of modes [except between the $(2N-2)\mathrm{th}$ and $2N\mathrm{th}$ modes].

%The squeezing correlation effect, especially the quantum entanglement emerges from the interplay of squeezing interactions and the topological edge states. The squeezing interactions build relationships between two quadratures. We note that the squeezing interactions themselves can generate quantum entanglement when the dissipation rate $\kappa$ is not much small than the coupling strengths, as we discussed in Sec. \ref{sec:two}. In this case, we can not make the approximation to neglect large eigenvalues, and there are stationary entanglements among all modes.

Therefore, as the absolute eigenenergies of the topological edge states are much smaller than those of the bulk states, we can selectly enhance the stationary entanglement between two topological edge states, when the dissipation rate $\kappa$ is much smaller than the coupling strengths.

\section{Topology-induced entanglement between two edge modes}\label{sec:tie}

As the topological edge states are most distributed at two edge modes, they have the maximal quantum entanglement. For $m$, $m^{\prime}\in\{1,N\}$, Eq. (\ref{eq:easy21})-(\ref{eq:easy31}) can be reduced to
\begin{equation}\label{eq:easy2}
	\langle \tilde{a}_{1}^{2}\rangle_{\mathrm{s}}=\langle \tilde{a}_{2N}^{2}\rangle_{\mathrm{s}}\approx\frac{\kappa^{2} l^{4}}{\kappa^{2}+4\delta^{2}}\left[e^{-2r_{0}}L_{1}-e^{2r_{0}}L_{2}\right],
\end{equation}
\begin{equation}\label{eq:easy3}
	\langle \tilde{a}_{1}\tilde{a}_{2N}\rangle_{\mathrm{s}}\approx-\frac{2i\delta\kappa l^{4}}{\kappa^{2}+4\delta^{2}}\left[e^{-2r_{0}}L_{1}-e^{2r_{0}}L_{2}\right].
\end{equation}

Moreover, in this case, we can neglect other terms except for $j=1,2$ in Eq. (\ref{eq:gg1}), because the topological edge states have distinct profiles compared with the bulk states. For $m$, $m^{\prime}\in\{1,N\}$, the distributions of the bulk states in Eq. (\ref{eq:gg1}) are much smaller than the edge states ($|\alpha_{j,m}\alpha_{j,m^{\prime}}|_{j=1,2}\gg|\alpha_{j,m}\alpha_{j,m^{\prime}}|_{j\neq1,2}$). Consequently, the summations can be reduced to $\langle \tilde{a}_{1}^{\dag}\tilde{a}_{2N}\rangle_{\mathrm{s}}=\langle\tilde{a}_{2N}^{\dag}\tilde{a}_{1}\rangle_{\mathrm{s}}=0$, and
\begin{equation}\label{eq:easy1}
	\langle \tilde{a}_{1}^{\dag}\tilde{a}_{1}\rangle_{\mathrm{s}}=\langle \tilde{a}_{2N}^{\dag}\tilde{a}_{2N}\rangle_{\mathrm{s}}\approx l^{4}\left(e^{-2r_{0}}L_{1}+e^{2r_{0}}L_{2}\right)-l^{2}.
\end{equation}

Due to the symmetry between two edge modes, the logarithmic negativity is given by $E_{N}=\mathrm{max}[0,-\ln2\eta^{-}]$, where $\eta^{-}=\left|\sqrt{(1/2+K_{1})^{2}-K_{2}^{2}}-K_{3}\right|$, and $K_{1}=\langle \tilde{a}_{1(2N)}^{\dag}\tilde{a}_{1(2N)}\rangle_{\mathrm{s}}$, $K_{2}=\langle \tilde{a}_{1(2N)}^{2}\rangle_{\mathrm{s}}$, $K_{3}=-i\langle \tilde{a}_{1}\tilde{a}_{2N}\rangle_{\mathrm{s}}$ are three different stationary mean values of the second-order moments. We note here we directly calculate the logarithmic negativity using the squeezed operators because the value of logarithmic negativity is independent of the squeezing transformation.

\begin{figure}[t]
	\centering
	\includegraphics[width=0.48\textwidth]{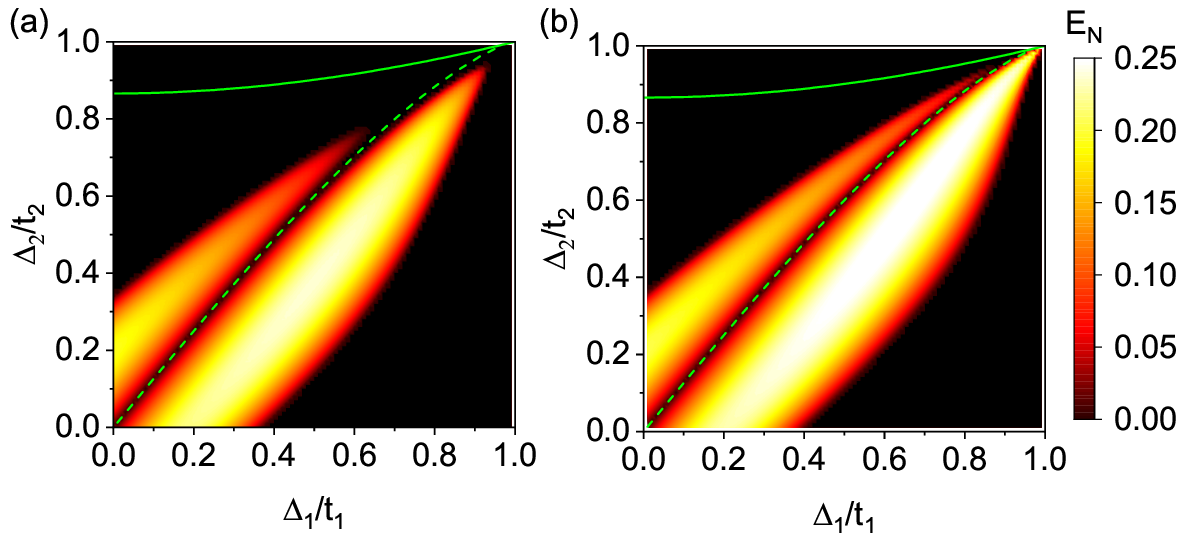}
	\caption{Stationary entanglement as quantified by the logarithmic negativity $E_{N}$ between two edge modes as functions of $\Delta_{1}/t_{1}$ and $\Delta_{2}/t_{2}$. (a) Exact numerical results obtained from the quantum master equations. (b) Approximate analytical results obtained from the quantum Langevin equations. The green solid line indicates the phase boundary between the trivial phase (above the line) and the topological phase (below the line).  The green dashed line indicates the vanished entanglement for $e^{-2r_{0}}L_{1}=e^{2r_{0}}L_{2}$. The parameters are $t_{2}/t_{1}=4$, $N=5$ and $\kappa=0.01$.}
	\label{fig:b}
\end{figure}

\begin{figure}[b]
	\centering
	\includegraphics[width=0.48\textwidth]{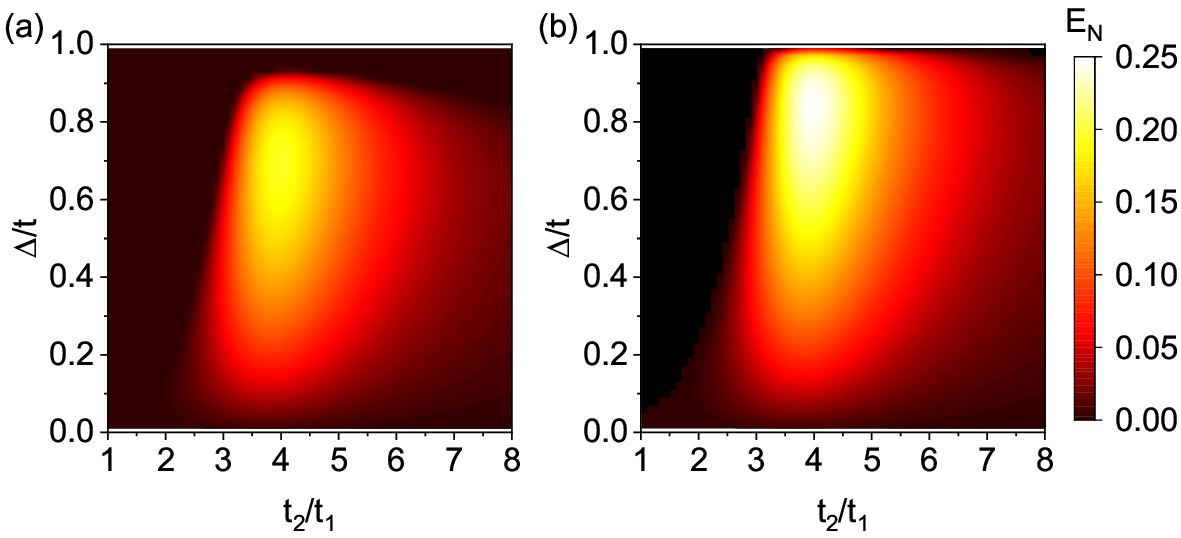}
	\caption{Stationary entanglement as quantified by the logarithmic negativity $E_{N}$ between two edge modes as functions of $t_{2}/t_{1}$ and $\Delta/t$ for $\Delta_{1}/t_{1}=\Delta_{2}/t_{2}$. (a) Exact numerical results obtained from the quantum master equations. (b) Approximate analytical results obtained from the quantum Langevin equations.
Other parameters are $N=5$ and $\kappa=0.01$.}
	\label{fig:c}
\end{figure}

To verify the above results, in Fig. \ref{fig:b} and Fig. \ref{fig:c} we plot the logarithmic negativity as functions of the system parameters for an open chain with 10 modes (N=5), where both exact numerical results and approximate analytical results are presented.
In Fig. \ref{fig:b}, the coupling strengths satisfy $t_{2}/t_{1}=4$, and thus the system is in the topological phase when $|\Delta_{2}/t_{2}|<\sqrt{3+(\Delta_{1}/t_{1})^{2}}/2$. The green solid line indicates the phase boundary between the topological phase (above the line) and the trivial phase (below the line). In Fig. \ref{fig:c}, the coupling strengths satisfy $\Delta_{1}/t_{1}=\Delta_{2}/t_{2}$, and thus the system is in the topological phase when $t_{2}/t_{1}>1$ (including all the region in Fig. \ref{fig:c}). We can find that for a wide parameter region in the topological phase, the logarithmic negativity is nonzero.
%Here we only consider real-valued coupling strengths. 

Remarkably, the approximate analytical results agree well with the exact numerical results obtained from the quantum master equations, which means that our approximations in the derivation of the analytical results perfectly catch the key point of the entanglement phenomenon. It is the existence of the topological edge states that leads to the stationary entanglement between two edge modes.

\section{Maximizing entanglement}\label{sec:max}

The analytical solutions can also help us to understand the pattern of the logarithmic negativity and to maximize the entanglement. In Fig. \ref{fig:b}, the logarithmic negativity splits into two bright areas. In the analytical expression, the dark area between the two bright areas corresponds to the case when $e^{-2r_{0}}L_{1}-e^{2r_{0}}L_{2}\approx0$ (denoted by the green dashed line). The factor $e^{-2r_{0}}L_{1}-e^{2r_{0}}L_{2}$ appears both in the squeezing term [cf. Eq. (\ref{eq:easy2})] and in the correlation term [cf. Eq. (\ref{eq:easy3})]. So when the entanglement disappears in the central dark area, the steady state of every mode is nearly an unsqueezed coherent state in the squeezing representation (but a squeezed state in the original representation), which is similar to the behaviors in the trivial phase. In this case, the entanglement is totally suppressed and there is only the single-mode squeezing effect.

\begin{figure}[t]
	\centering
	\includegraphics[width=0.48\textwidth]{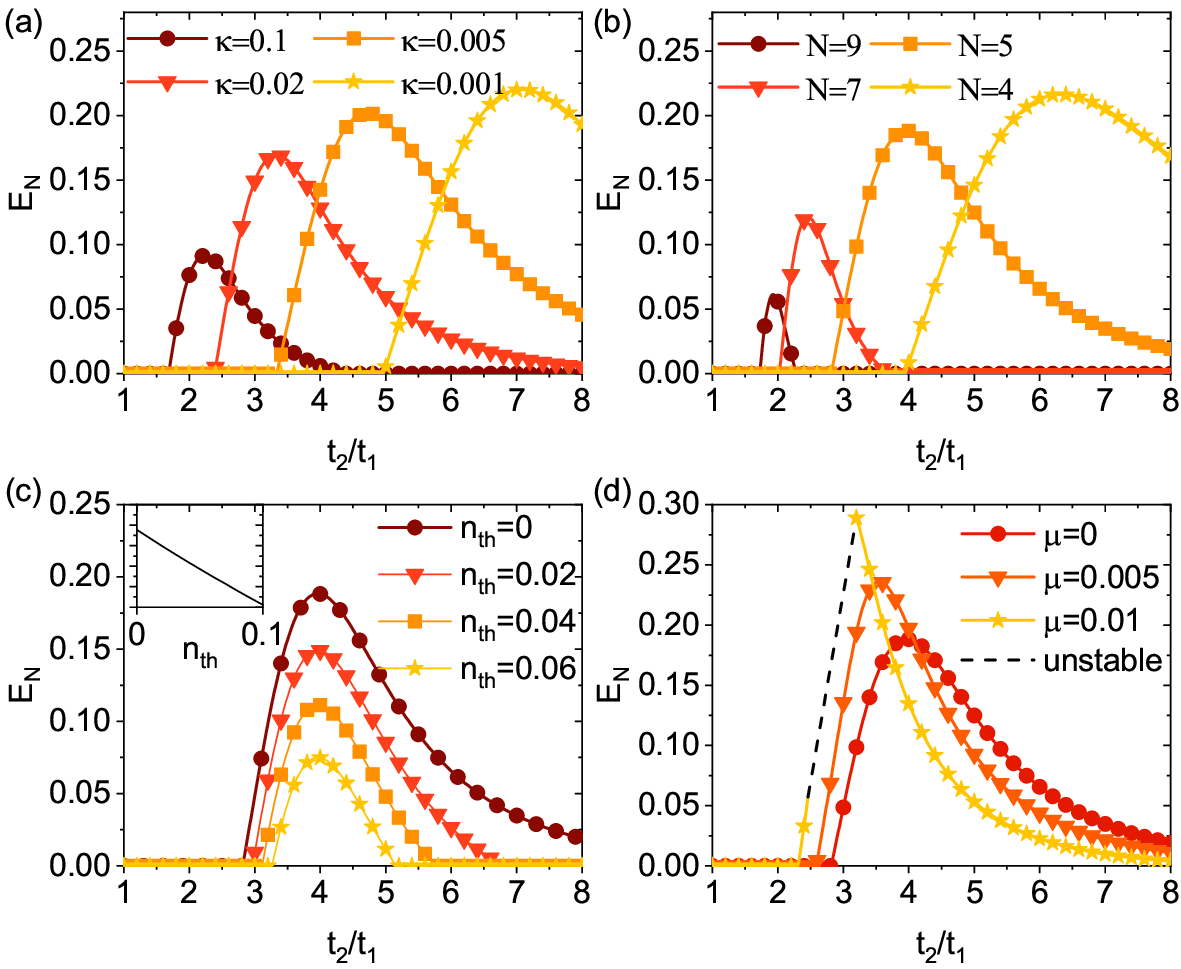}
	\caption{Stationary entanglement between two edge modes (quantified by $E_{N}$) versus the ratio of linear coupling strengths $t_{2}/t_{1}$ considering the influence of the dissipation rate $\kappa$ (a), the number of the unit cell $N$ (b), the environment photon number $n_{\mathrm{th}}$ (c) and the chemical potential $\mu$ (d). The squeezing interactions satisfy $\Delta_{1}/t_{1}=\Delta_{2}/t_{2}=0.6$. The inset of panel (c) is the logarithmic negativity $E_{N}$ as a function of $n_{\mathrm{th}}$ for $t_{2}/t_{1}=4$. The dashed line in (d) denotes the region where the system is unstable (no stationary solutions). Other parameters are: $\mu=0$ in (a), (b), and (c); $N=5$ in (a), (c), and (d); $n_{\mathrm{th}}=0$ in (a), (b), and (d); $\kappa=0.01$ in (b), (c), and (d).}
	\label{fig:d}
\end{figure}

We then focus on the special case $\Delta_{1}/t_{1}=\Delta_{2}/t_{2}$ considered in Fig. \ref{fig:c}. It is the condition when the non-Hermitian skin effect disappears $(\Sigma_{1}\Sigma_{2}=\Sigma_{3}^{2})$. In this case, the squeezing parameters become $r_{\mathrm{a}}=r_{\mathrm{b}}=2r_{0}$, and the mean values of the second-order moments [cf. Eq. (\ref{eq:easy2})-(\ref{eq:easy1})] can be greatly reduced, which are
\begin{equation}\label{eq:ea1}
	\langle \tilde{a}_{1}^{\dag}\tilde{a}_{1}\rangle_{\mathrm{s}}=\langle \tilde{a}_{2N}^{\dag}\tilde{a}_{2N}\rangle_{\mathrm{s}}\approx l^{2}\frac{e^{-2r_{0}}+e^{2r_{0}}-2}{2},
\end{equation}
\begin{equation}\label{eq:ea2}
	\langle \tilde{a}_{1}^{2}\rangle_{\mathrm{s}}=\langle \tilde{a}_{2N}^{2}\rangle_{\mathrm{s}}\approx\frac{\kappa^{2} l^{2}}{\kappa^{2}+4\delta^{2}}\frac{e^{-2r_{0}}-e^{2r_{0}}}{2},
\end{equation}
\begin{equation}\label{eq:ea3}
	\langle \tilde{a}_{1}\tilde{a}_{2N}\rangle_{\mathrm{s}}\approx-\frac{2i\delta\kappa l^{2}}{\kappa^{2}+4\delta^{2}}\frac{e^{-2r_{0}}-e^{2r_{0}}}{2}.
\end{equation}
Then the logarithmic negativity mainly depends on the interplay between the dissipation $\kappa$ and the absolute energy of the topological edge modes $\delta$. According to Eq. (\ref{eq:ea2})-(\ref{eq:ea3}), the maximal logarithmic negativity is obtained near $\kappa=2\delta$. We note that the absolute energy of the topological edge modes $\delta$ is smaller when the ratio $t_{2}/t_{1}$ is larger. Inversely, the absolute energy $\delta$ is smaller when the number of unit cells $N$ is smaller. So for a smaller dissipation rate or a smaller unit cell number, the maximal entanglement is obtained at a larger ratio $t_{2}/t_{1}$, as shown in Fig. \ref{fig:d}(a) and \ref{fig:d}(b).
% In Fig. \ref{fig:d}, we plot the logarithmic negativity as a function of the ratio of linear coupling strengths considering the influence of the dissipation rate $\kappa$, the number of the unit cells $N$ and the chemical potential $\mu$.

We also investigate the influence of the environment photon number $n_{\mathrm{th}}$ and the chemical potential (on-site energy) $\mu$ on the stationary entanglement, which are not included in the above calculation. For simplicity, we assume $n_{\mathrm{th}}$ and $\mu$ of all modes are the same. 
As plotted in Fig. \ref{fig:d}(c), the stationary entanglement decays linearly when increasing the environment photon number $n_{\mathrm{th}}$. As shown in Fig. \ref{fig:d}(d), the chemical potential will also affect the entanglement. In some parameter ranges when the system is in the stable region, the chemical potential can enhance the maximal entanglement, while it can also enhances the instability due to the intrinsic non-Hermiticity of the squeezing interactions.

\section{Stationary entanglements with complex-valued couplings}\label{sec:complex}

\begin{figure}[t]
	\centering
	\includegraphics[width=0.4\textwidth]{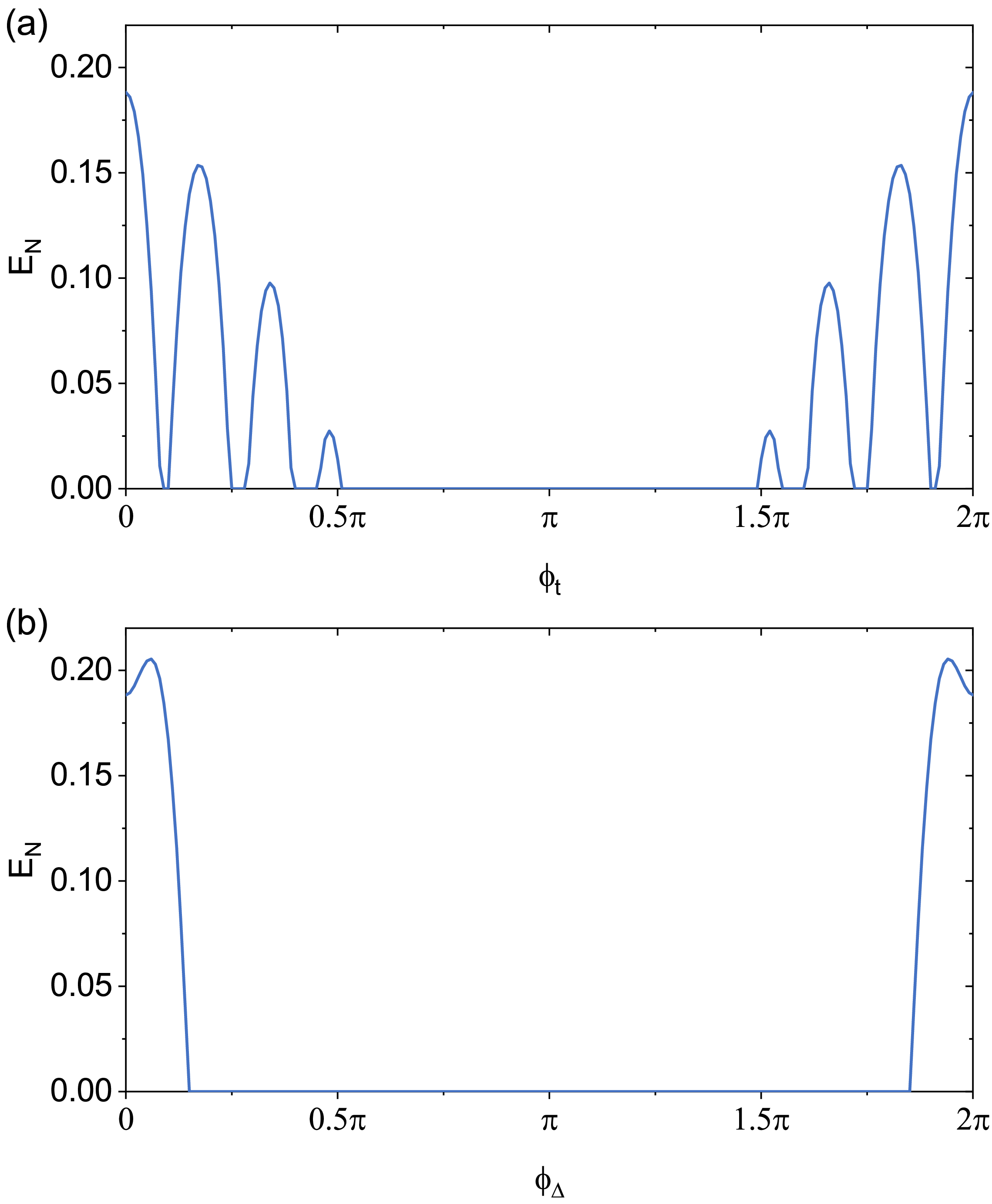}
	\caption{Stationary entanglement (quantified by $E_{N}$) between two edge modes versus the coupling phase $\phi_{\mathrm{t}}$ (a) and $\phi_{\Delta}$ (b). Other parameters are $\kappa=0.01$, $N=5$, $\mu=0$, $t_{1}=1$, $|t_{2}|=4$ and $\Delta_{1}/t_{1}=|\Delta_{2}/t_{2}|=0.6$.}
	\label{fig:6}
\end{figure}

As shown in Sec. \ref{sec:topo}, the topological phase transition is independent of the coupling phases. However, the stationary entanglements are highly dependent on the coupling phases. This is because the squeezing transformation is phase-dependent. 
%The main influence of the coupling phases is that the squeezing transformation becomes different and far more complex. So we only obtain numerical results in this case.
Figure \ref{fig:6} plots the stationary entanglement (quantified by $E_{N}$) between two edge modes versus the coupling phase $\phi_{\mathrm{t}}$ and $\phi_{\Delta}$. The stationary entanglement exhibits an interesting finger-like pattern versus the coupling phase $\phi_{\mathrm{t}}$, while there is stationary entanglement only for a small range of coupling phase $\phi_{\Delta}$ near $0$ or $2\pi$.

In particular, the case for $\pi$ phase can be understood through the analytical expression, as the coupling strengths are still real. The cases for $\phi_{\mathrm{t}}=\pi$ and $\phi_{\Delta}=\pi$ are equivalent. For example, in the case of $\phi_{\Delta}=\pi$, the squeezing parameters $r_{\mathrm{b}}$ in the squeezing transformation becomes negative, while $r_{\mathrm{a}}$ is still positive. Then the squeezing parameter of every mode $r_{j}$ is greatly enhanced, which leads to more enhancement of the stationary photon number Eq. (\ref{eq:easy1}) than the enhancement of the squeezing correlation Eq. (\ref{eq:easy2}), so the quantum entanglement disappears.

\section{Experimental realization}\label{sec:exp}

The main requirements of the system are site-dependent coupling strengths and the squeezing interactions. These requirements are already satisfied by a recent experiment based on an optomechanical cavity \cite{del_pino_non-hermitian_2022}. They make use of the idea of synthetic dimension realized from multiple non-degenerate mechanical modes. These mechanical modes are coupled to an optical cavity mode through the radiation pressure, and the optical cavity mode can be used to generate both the beamsplitter (linear) and squeezing interactions between different mechanical modes. These couplings are obtained through modulation at a special frequency in the large-detuning regime. Moreover, the coupling strengths can be individually controlled by the modulation depth. So the multimode optomechanical system is a perfect platform to realize the topology-induced entanglement, and the stationary entanglement can be read out by an additional probe laser. As shown in Fig. \ref{fig:d}(b), there are obvious quantum entanglements for only four modes. 
%A major difficulty is that this quantum entanglement is sensitive to the environment temperature. In principle, it can be solved by employing optomechanical cooling to all the mechanical modes.

\section{Discussion and Conclusion}\label{sec:sum}

We establish a direct relationship between quantum entanglement and classical topology. It is distinct from the proposals utilizing topology to enhance the robustness of quantum effects \cite{rechtsman_topological_2016,blanco-redondo_topological_2018,wang_topologically_2019-1,tschernig_topological_2021,wang_topological_2019,wang_topologically_2019,dai_topologically_2022,ren_topologically_2022}. It is also different from the efforts to include quantum effects to obtain novel topological phase transition \cite{peano_topological_2016,cai_topological_2021,deng_observing_2022}. Our work shows that the bosonic topology can be a source of the quantum entanglements and the quantum entanglements can be a quantum signature of the topological phase. It also has the potential to investigate quantum phase transition driven by bosonic topology.

The results in this work reveal a general mechanism that can be applied to various systems and can be generalized to higher dimensions. For example, this mechanism can be directly applied to the lattice model with dissipative pairing interactions \cite{pocklington_dissipative_2023} and the model with single-mode squeezing \cite{peano_topological_2016}. Moreover, this mechanism can be generalized to high-dimensional systems such as the higher-order topological corner modes, and the mechanism can also be used to generate quantum entanglements as a witness of the Floquet topology.

In summary, we discover that there is topology-induced entanglement effect in the steady state of a bosonic quadratic chain. We show the stationary entanglement only exists in the topological phase.
The relation between the entanglement and the topological edge states is established with analytical expressions by appropriately solving the quantum Langevin equations, where we neglect the terms containing bulk-state eigenenergies but keep the terms containing near-zero eigenenergies which correspond to the topological edge states.
The analytical results show good agreement with the numerical results obtained from the covariance approach based on the quantum master equations, which proves that our approximation perfectly catches the key point of the emerging entanglement phenomenon.
We verify that the approximation is valid because the squeezing correlations are greatly suppressed when the intrasystem coupling strengths (which determine the system eigenenergies) do not match the system-environment coupling strengths (denoted by the dissipation rate).
For a topological system, the topological edge states possess near-zero eigenenergies, which are much smaller than the absolute value of the eigenenergies of the bulk states, so we can selectively match the topological edge states with the system-environment coupling and generate obvious stationary entanglements between these states. This kind of topological matching and related entanglements disappears in the trivial phase when there are no topological edge states. Based on this finding, we thoroughly discuss the influence of different parameters on the stationary entanglements and maximal conditions.
%The stationary entanglements mainly exist between edge modes due to the localized profiles of the topological edge states.
This model is implementable in a variety of experimental platforms, such as multimode optomechanical systems and superconducting quantum circuits. Our work opens an avenue for investigating quantum entanglement in topological systems.

\begin{acknowledgments}
	This work is supported by the Key-Area Research and Development Program of Guangdong Province (Grant No. 2019B030330001), the National Natural Science Foundation of China (NSFC) (Grant Nos. 12275145, 92050110, 91736106, 11674390, and 91836302), and the National Key R\&D Program of China (Grants No. 2018YFA0306504).
\end{acknowledgments}

\appendix

\begin{widetext}
\section{Bloch theory for a quadratic chain}\label{ap:b}

In this section, we provide a detailed calculation of the Bloch theory for a quadratic chain with staggered couplings. The Hamiltonian is written as
\begin{equation}\label{eq:sh1}
	\begin{split}
		H_{\mathrm{}}=&\sum_{j=1}^{N}(t_{\mathrm{}1}a_{2j-1}^{\dag} a_{2j}+\Delta_{\mathrm{}1} a_{2j-1}^{\dag} a_{2j}^{\dag}+\mathrm{H.c.})\\
		+&\sum_{j=1}^{N-1}(t_{\mathrm{}2} a_{2j+1}^{\dag} a_{2j}+\Delta_{\mathrm{}2} a_{2j+1}^{\dag} a_{2j}^{\dag}+\mathrm{H.c.}),
	\end{split}
\end{equation}
where $t_{\mathrm{}1}$ ($t_{\mathrm{}2}$) and $\Delta_{\mathrm{}1}$ ($\Delta_{\mathrm{}2}$) are the intracell (intercell) coupling strengths of linear and squeezing interactions, respectively, $N$ is the number of unit cells, and $a_{j}$ is the annihilation operator of the $j$th mode. After the Fourier transformation, the Bloch Hamiltonian of the system can be written as
\begin{equation}
	\mathcal{H}_{\mathrm{}}(k)=(t_{\mathrm{}1}+t_{\mathrm{}2}e^{ik})a_{k}^{\dag}a_{k}^{\prime}+(\Delta_{\mathrm{}1}+\Delta_{\mathrm{\mathrm{}2}}e^{ik})a_{k}^{\dag}a_{-k}^{\prime\dag}+\mathrm{H.c.},
\end{equation}
or in the matrix form: $\mathcal{H}(k)=\frac{1}{2}C_{k}^{\dag}\mathcal{H}_{\mathrm{M}}(k)C_{k}$, where $C_{k}^{\dag}=(a_{k}^{\dag},a_{k}^{\prime\dag},a_{-k},a_{-k}^{\prime})$ and
\begin{equation}
	\mathcal{H}_{\mathrm{M}}(k)=\begin{pmatrix}
		0                                                      & t_{\mathrm{1}}+t_{\mathrm{2}}e^{ik}                   & 0                                              & \Delta_{\mathrm{1}}+\Delta_{\mathrm{2}}e^{ik} \\
		t_{\mathrm{1}}^{*}+t_{\mathrm{2}}^{*}e^{-ik}           & 0                                                     & \Delta_{\mathrm{1}}+\Delta_{\mathrm{2}}e^{-ik} & 0                                             \\
		0                                                      & \Delta_{\mathrm{1}}^{*}+\Delta_{\mathrm{2}}^{*}e^{ik} & 0                                              & t_{\mathrm{1}}^{*}+t_{\mathrm{2}}^{*}e^{ik}   \\
		\Delta_{\mathrm{1}}^{*}+\Delta_{\mathrm{2}}^{*}e^{-ik} & 0                                                     & t_{\mathrm{1}}+t_{\mathrm{2}}e^{-ik}           & 0
	\end{pmatrix}.
\end{equation}
Here the excitation modes are Bogoliubov modes that are determined by the eigenvalue equation of $\tau_{z}\mathcal{H}_{\mathrm{M}}$, where $\tau_{z}=\mathrm{Diagonal}(\mathbbm{1},-\mathbbm{1})$ and $\mathbbm{1}$ is an identity matrix with half the dimension of the corresponding Hamiltonian \cite{yokomizo_non-bloch_2021}. The eigenvalue equation $\mathrm{det}|\tau_{z}\mathcal{H}_{\mathrm{M}}-\xi \mathbbm{1}|=0$ can be obtained as
\begin{equation}\label{eq:eigen1}
	\begin{split}
		&\left(\xi^{2}-|t_{\mathrm{1}}+t_{\mathrm{2}}e^{ik}|^{2}\right)\left(\xi^{2}-|t_{\mathrm{1}}+t_{\mathrm{2}}e^{-ik}|^{2}\right)+\xi^{2}\left(|\Delta_{\mathrm{1}}+\Delta_{\mathrm{2}}e^{ik}|^{2}+|\Delta_{\mathrm{1}}+\Delta_{\mathrm{2}}e^{-ik}|^{2}\right)+\\
		|(\Delta_{\mathrm{1}}&+\Delta_{2}e^{ik})(\Delta_{\mathrm{1}}+\Delta_{2}e^{-ik})|^{2}-2\mathrm{Re}\left[(t_{\mathrm{1}}+t_{\mathrm{2}}e^{ik})(t_{\mathrm{1}}^{*}+t_{\mathrm{2}}^{*}e^{ik})(\Delta_{\mathrm{1}}+\Delta_{\mathrm{2}}e^{-ik})(\Delta_{\mathrm{1}}^{*}+\Delta_{\mathrm{2}}^{*}e^{-ik})\right]=0.
	\end{split}
\end{equation}
% \end{widetext}
Consequently, only the relative phases between $t_{1}$ ($\Delta_{1}$) and $t_{2}$ ($\Delta_{2}$) are important. So we can assume
\begin{equation}
	t_{\mathrm{2}}=q_{\mathrm{t}}t_{\mathrm{1}}e^{i\phi_{\mathrm{t}}},\Delta_{\mathrm{2}}=q_{\mathrm{\Delta}}\Delta_{\mathrm{1}}e^{i\phi_{\Delta}},
\end{equation}
where $q_{\mathrm{t},\Delta}$, $t_{\mathrm{1}}$ and $\Delta_{\mathrm{1}}$ are all real, $\phi_{\mathrm{t}}$ ($\phi_{\Delta}$) are the relative phases between $t_{1}$ ($\Delta_{1}$) and $t_{2}$ ($\Delta_{2}$), respectively. So Eq. (\ref{eq:eigen1}) can be rewritten as
% \begin{widetext}
\begin{equation}
	\begin{split}
		&\xi^{4}-2\xi^{2}\left[(1+q_{\mathrm{t}}^{2}+2q_{\mathrm{t}}\cos k\cos \phi_{\mathrm{t}})t_{\mathrm{1}}^{2}-(1+q_{\Delta}^{2}+2q_{\Delta}\cos k\cos \phi_{\Delta})\Delta_{\mathrm{1}}^{2}\right]\\
		+&\left[1+q_{\mathrm{t}}^{2}+2q_{\mathrm{t}}\cos (k+\phi_{\mathrm{t}})\right]\left[1+q_{\mathrm{t}}^{2}+2q_{\mathrm{t}}\cos (k-\phi_{\mathrm{t}})\right]t_{\mathrm{1}}^{4}\\
		+&\left[1+q_{\Delta}^{2}+2q_{\Delta}\cos (k+\phi_{\Delta})\right]\left[1+q_{\Delta}^{2}+2q_{\Delta}\cos (k-\phi_{\Delta})\right]\Delta_{\mathrm{1}}^{4}\\
		-&2\mathrm{Re}\left[(1+2q_{\mathrm{t}}e^{ik}\cos \phi_{\mathrm{t}}+q_{\mathrm{t}}^{2}e^{2ik})(1+2q_{\Delta}e^{-ik}\cos \phi_{\Delta}+q_{\Delta}^{2}e^{-2ik})\right]t_{\mathrm{1}}^{2}\Delta_{\mathrm{1}}^{2}=0.
	\end{split}
\end{equation}
Then we can obtain
\begin{equation}\label{eq:slambda1}
	\begin{split}
		\xi^{2}=&(1+q_{\mathrm{t}}^{2}+2q_{\mathrm{t}}\cos k\cos \phi_{\mathrm{t}})t_{\mathrm{1}}^{2}-(1+q_{\Delta}^{2}+2q_{\Delta}\cos k\cos \phi_{\Delta})\Delta_{\mathrm{1}}^{2}\\
		&\pm2|\sin k|\sqrt{(q_{\mathrm{t}}\sin\phi_{\mathrm{t}}t_{\mathrm{1}}^{2})^{2}+(q_{\Delta}\sin\phi_{\Delta}\Delta_{\mathrm{1}}^{2})^{2}-t_{\mathrm{1}}^{2}\Delta_{\mathrm{1}}^{2}\left(q_{\mathrm{t}}^{2}+q_{\Delta}^{2}-2q_{\mathrm{t}}q_{\Delta}\cos \phi_{\mathrm{t}}\cos \phi_{\Delta}\right)},
	\end{split}
\end{equation}
or
\begin{equation}\label{eq:slambda3}
	\begin{split}
		\xi^{2}=&(1+q_{\mathrm{t}}^{2}+2q_{\mathrm{t}}\cos k\cos \phi_{\mathrm{t}})t_{\mathrm{1}}^{2}-(1+q_{\Delta}^{2}+2q_{\Delta}\cos k\cos \phi_{\Delta})\Delta_{\mathrm{1}}^{2}\\
		&\pm2|\sin k|\sqrt{(q_{\mathrm{t}}^{2}t_{\mathrm{1}}^{2}-q_{\Delta}^{2}\Delta_{\mathrm{1}}^{2})(t_{\mathrm{1}}^{2}-\Delta_{\mathrm{1}}^{2})-(q_{\mathrm{t}}\cos \phi_{\mathrm{t}}t_{\mathrm{1}}^{2}-q_{\Delta}\cos \phi_{\Delta}\Delta_{\mathrm{1}}^{2})^{2}},
	\end{split}
\end{equation}
For simplicity, we let $\Sigma_{1}=t_{\mathrm{}1}^{2}-\Delta_{\mathrm{}1}^{2}$, $\Sigma_{2}=q_{\mathrm{t}}^{2}t_{\mathrm{}1}^{2}-q_{\Delta}^{2}\Delta_{\mathrm{}1}^{2}$ and $\Sigma_{3}=q_{\mathrm{t}}\cos \phi_{\mathrm{t}}t_{\mathrm{}1}^{2}-q_{\Delta}\cos \phi_{\Delta}\Delta_{\mathrm{}1}^{2}$. So Eq. (\ref{eq:slambda3}) becomes
\begin{equation}\label{eq:slambda}
	\xi^{2}=\Sigma_{1}+\Sigma_{2}+2\Sigma_{3}\cos k\pm 2|\sin k|\sqrt{\Sigma_{1}\Sigma_{2}-\Sigma_{3}^{2}},
\end{equation}
corresponding to Eq. (3) in the main text. When $\Sigma_{1}\Sigma_{2}-\Sigma_{2}^{2}>0$, the system does not exhibit the non-Hermitian skin effect. In this case, Eq. (\ref{eq:slambda}) can be rewritten as
\begin{equation}
	\xi^{2}=\Sigma_{1}+\Sigma_{2}+2\Sigma_{1}\Sigma_{2}\cos (\pm k+\varphi),
\end{equation}
where $\tan \varphi=\sqrt{\Sigma_{1}\Sigma_{2}-\Sigma_{3}^{2}}/\Sigma_{3}$. Then we can obtain the energy spectrum as
\begin{equation}\label{eq:sband}
	(\sqrt{\Sigma_{1}}-\sqrt{\Sigma_{2}})^{2}<\xi^{2}<(\sqrt{\Sigma_{1}}+\sqrt{\Sigma_{2}})^{2},
\end{equation}
where we assume $\Sigma_{1,2}$ are both positive.

\section{Non-Bloch theory for a quadratic chain}\label{ap:nb}

When there is the non-Hermitian skin effect, i.e., $\Sigma_{1}\Sigma_{2}-\Sigma_{3}^{2}<0$, the Bloch theory fails in the calculation of the open-boundary bulk spectrum. Then we need to use the non-Bloch theory, with the replacements $e^{ik}\to\beta$ and $e^{-ik}\to\beta^{-1}$. So the non-Bloch Hamiltonian matrix can be written as
\begin{equation}
	\mathcal{H}_{\mathrm{M}}(\beta)=\begin{pmatrix}
		0                                                         & t_{\mathrm{1}}+t_{\mathrm{2}}\beta                   & 0                                                 & \Delta_{\mathrm{1}}+\Delta_{\mathrm{2}}\beta \\
		t_{\mathrm{1}}^{*}+t_{\mathrm{2}}^{*}\beta^{-1}           & 0                                                    & \Delta_{\mathrm{1}}+\Delta_{\mathrm{2}}\beta^{-1} & 0                                            \\
		0                                                         & \Delta_{\mathrm{1}}^{*}+\Delta_{\mathrm{2}}^{*}\beta & 0                                                 & t_{\mathrm{1}}^{*}+t_{\mathrm{2}}^{*}\beta   \\
		\Delta_{\mathrm{1}}^{*}+\Delta_{\mathrm{2}}^{*}\beta^{-1} & 0                                                    & t_{\mathrm{1}}+t_{\mathrm{2}}\beta^{-1}           & 0
	\end{pmatrix}.
\end{equation}
The eigenvalue equation $\mathrm{det}|\tau_{z}\mathcal{H}_{\mathrm{M}}-\xi \mathbbm{1}|=0$ is
\begin{equation}
	\begin{split}
		&\xi^{4}-2\xi^{2}\left[(1+q_{\mathrm{t}}^{2}+q_{\mathrm{t}}\cos \phi_{\mathrm{t}}(\beta+\beta^{-1}))t_{\mathrm{1}}^{2}-(1+q_{\Delta}^{2}+q_{\Delta}\cos \phi_{\Delta}(\beta+\beta^{-1}))\Delta_{\mathrm{1}}^{2}\right]\\
		+&\left\{\left[1+q_{\mathrm{t}}^{2}+q_{\mathrm{t}}\cos \phi_{\mathrm{t}}(\beta+\beta^{-1})\right]^{2}+\left[q_{\mathrm{t}}\sin \phi_{\mathrm{t}}(\beta-\beta^{-1})\right]^{2}\right\}t_{\mathrm{1}}^{4}\\
		+&\left\{\left[1+q_{\Delta}^{2}+q_{\Delta}\cos \phi_{\Delta}(\beta+\beta^{-1})\right]^{2}+\left[q_{\Delta}\sin \phi_{\Delta}(\beta-\beta^{-1})\right]^{2}\right\}\Delta_{\mathrm{1}}^{4}\\
		-&\left[2(1+q_{\mathrm{t}}^{2}q_{\Delta}^{2})+(q_{\mathrm{t}}^{2}+q_{\Delta}^{2})(\beta^{2}+\beta^{-2})+8\cos\phi_{\mathrm{t}}\cos\phi_{\Delta}q_{\mathrm{t}}q_{\Delta}\right]t_{\mathrm{1}}^{2}\Delta_{\mathrm{1}}^{2}\\
		-&2\left[\cos\phi_{\mathrm{t}}q_{\mathrm{t}}(1+q_{\Delta}^{2})+\cos\phi_{\Delta}q_{\Delta}(1+q_{\mathrm{t}}^{2})\right](\beta+\beta^{-1})t_{\mathrm{1}}^{2}\Delta_{\mathrm{1}}^{2}=0.
	\end{split}
\end{equation}
Then we obtain
\begin{equation}\label{eq:nblambda1}
	\begin{split}
		\xi^{2}=&[1+q_{\mathrm{t}}^{2}+q_{\mathrm{t}}\cos \phi_{\mathrm{t}}(\beta+\beta^{-1})]t_{\mathrm{1}}^{2}-[1+q_{\Delta}^{2}+q_{\Delta}\cos \phi_{\Delta}(\beta+\beta^{-1})]\Delta_{\mathrm{1}}^{2}\\
		&\pm\sqrt{-(\beta-\beta^{-1})^{2}}\sqrt{(q_{\mathrm{t}}\sin\phi_{\mathrm{t}}t_{\mathrm{1}}^{2})^{2}+(q_{\Delta}\sin\phi_{\Delta}\Delta_{\mathrm{1}}^{2})^{2}-t_{\mathrm{1}}^{2}\Delta_{\mathrm{1}}^{2}\left(q_{\mathrm{t}}^{2}+q_{\Delta}^{2}-2q_{\mathrm{t}}q_{\Delta}\cos \phi_{\mathrm{t}}\cos \phi_{\Delta}\right)},
	\end{split}
\end{equation}
or
\begin{equation}\label{eq:nblambda3}
	\begin{split}
		\xi^{2}=&[1+q_{\mathrm{t}}^{2}+q_{\mathrm{t}}\cos \phi_{\mathrm{t}}(\beta+\beta^{-1})]t_{\mathrm{1}}^{2}-[1+q_{\Delta}^{2}+q_{\Delta}\cos \phi_{\Delta}(\beta+\beta^{-1})]\Delta_{\mathrm{1}}^{2}\\
		&\pm\sqrt{-(\beta-\beta^{-1})^{2}}\sqrt{(q_{\mathrm{t}}^{2}t_{\mathrm{1}}^{2}-q_{\Delta}^{2}\Delta_{\mathrm{1}}^{2})(t_{\mathrm{1}}^{2}-\Delta_{\mathrm{1}}^{2})-(q_{\mathrm{t}}\cos \phi_{\mathrm{t}}t_{\mathrm{v}}^{2}-q_{\Delta}\cos \phi_{\Delta}\Delta_{\mathrm{1}}^{2})^{2}}.
	\end{split}
\end{equation}
We can also obtain the generalized momentum as
\begin{equation}\label{eq:sbeta}
	\beta=\frac{1}{2}\frac{\lambda^{2}-(1+q_{\mathrm{t}}^{2})t_{\mathrm{1}}^{2}+(1+q_{\Delta}^{2})\Delta_{\mathrm{1}}^{2}\pm\sqrt{(\xi^{2}-(1+q_{\mathrm{t}}^{2})t_{\mathrm{1}}^{2}+(1+q_{\Delta}^{2})\Delta_{\mathrm{1}}^{2})^{2}-4(q_{\mathrm{t}}^{2}t_{\mathrm{1}}^{2}-q_{\Delta}^{2}\Delta_{\mathrm{1}}^{2})(t_{\mathrm{1}}^{2}-\Delta_{\mathrm{1}}^{2})}}{q_{\mathrm{t}}\cos \phi_{\mathrm{t}}t_{\mathrm{1}}^{2}-q_{\Delta}\cos \phi_{\Delta}\Delta_{\mathrm{1}}^{2}\pm \sqrt{(q_{\mathrm{t}}\cos \phi_{\mathrm{t}}t_{\mathrm{1}}^{2}-q_{\Delta}\cos \phi_{\Delta}\Delta_{\mathrm{1}}^{2})^{2}-(q_{\mathrm{t}}^{2}t_{\mathrm{1}}^{2}-q_{\Delta}^{2}\Delta_{\mathrm{1}}^{2})(t_{\mathrm{1}}^{2}-\Delta_{\mathrm{1}}^{2})}}.
\end{equation}
There are two ``$\pm$'' and four $\beta$. The four $\beta$ are two pairs according to the $\pm$ in the denominator. We note the denominator is real as the term under the root sign is positive. The existence of the generalized Brillouin zone requires the absolute values of two $\beta$ in each pair equal to each other. To be clear, we let
\begin{equation}\label{eq:sbeta2}
	\beta_{i,\pm}=\frac{1}{2}\frac{\lambda^{2}-(1+q_{\mathrm{t}}^{2})t_{\mathrm{1}}^{2}+(1+q_{\Delta}^{2})\Delta_{\mathrm{1}}^{2}\pm\sqrt{(\xi^{2}-(1+q_{\mathrm{t}}^{2})t_{\mathrm{1}}^{2}+(1+q_{\Delta}^{2})\Delta_{\mathrm{1}}^{2})^{2}-4(q_{\mathrm{t}}^{2}t_{\mathrm{1}}^{2}-q_{\Delta}^{2}\Delta_{\mathrm{1}}^{2})(t_{\mathrm{1}}^{2}-\Delta_{\mathrm{1}}^{2})}}{q_{\mathrm{t}}\cos \phi_{\mathrm{t}}t_{\mathrm{1}}^{2}-q_{\Delta}\cos \phi_{\Delta}\Delta_{\mathrm{1}}^{2}+(-1)^{i} \sqrt{(q_{\mathrm{t}}\cos \phi_{\mathrm{t}}t_{\mathrm{1}}^{2}-q_{\Delta}\cos \phi_{\Delta}\Delta_{\mathrm{1}}^{2})^{2}-(q_{\mathrm{t}}^{2}t_{\mathrm{1}}^{2}-q_{\Delta}^{2}\Delta_{\mathrm{1}}^{2})(t_{\mathrm{1}}^{2}-\Delta_{\mathrm{1}}^{2})}},
\end{equation}
for $i=1,2$. The requirement becomes $|\beta_{i,+}|=|\beta_{i,-}|$, which means the term under the root sign in the numerator is negative, i.e.,
\begin{equation}
	(\xi^{2}-(1+q_{\mathrm{t}}^{2})t_{\mathrm{1}}^{2}+(1+q_{\Delta}^{2})\Delta_{\mathrm{1}}^{2})^{2}-4(q_{\mathrm{t}}^{2}t_{\mathrm{1}}^{2}-q_{\Delta}^{2}\Delta_{\mathrm{1}}^{2})(t_{\mathrm{1}}^{2}-\Delta_{\mathrm{1}}^{2})<0.
\end{equation}
So we can obtain the energy spectrum as
\begin{equation}\label{eq:band2}
	\left(\sqrt{t_{\mathrm{1}}^{2}-\Delta_{\mathrm{1}}^{2}}-\sqrt{q_{\mathrm{t}}^{2}t_{\mathrm{1}}^{2}-q_{\Delta}^{2}\Delta_{\mathrm{1}}^{2}}\right)^{2}<\lambda^{2}<\left(\sqrt{t_{\mathrm{1}}^{2}-\Delta_{\mathrm{1}}^{2}}+\sqrt{q_{\mathrm{t}}^{2}t_{\mathrm{1}}^{2}-q_{\Delta}^{2}\Delta_{\mathrm{1}}^{2}}\right)^{2},
\end{equation}
which is the same as the energy spectrum Eq. (\ref{eq:band}) for $\Sigma_{1}\Sigma_{2}-\Sigma_{3}^{2}>0$ when there is no non-Hermitian skin effect.

\section{Derivation of the quantum Langevin equations}\label{ap:qle}

We start from the quantum Langevin equations of operators after the squeezing transformation [Eq. (\ref{eq:langevin1})-(\ref{eq:langevin2})], which are
\begin{equation}\label{eq:slangevin1}
	\dot{\tilde{a}}_{2j-1}=-\frac{\kappa}{2}\tilde{a}_{2j-1}-it_{1}^{\prime}\tilde{a}_{2j}-it_{2}^{\prime}\tilde{a}_{2j-2}-\sqrt{\kappa}\tilde{a}_{\mathrm{in},2j-1},
\end{equation}
\begin{equation}\label{eq:slangevin2}
	\dot{\tilde{a}}_{2j}=-\frac{\kappa}{2}\tilde{a}_{2j}-it_{1}^{\prime}\tilde{a}_{2j-1}-it_{2}^{\prime}\tilde{a}_{2j+1}-\sqrt{\kappa}\tilde{a}_{\mathrm{in},2j},
\end{equation}
where $\tilde{a}_{\mathrm{in},j}$ are the noise operators. Due to the squeezing transformation, these noise operators denote couplings to a squeezed environment. The above Langevin equations can be rewritten in the matrix form as
\begin{equation}
	\dot{\tilde{\bm{\mathrm{A}}}}=(-\frac{\kappa}{2}\mathbbm{1}-iS)\tilde{\bm{\mathrm{A}}}-\sqrt{\kappa}\tilde{\bm{\mathrm{A}}}_{\mathrm{in}},
\end{equation}
where $\mathbbm{1}$ is the identity matrix, $\tilde{\bm{\mathrm{A}}}=(\tilde{a}_{1},\cdots)^{\mathrm{T}}$, $\tilde{\bm{\mathrm{A}}}_{\mathrm{in}}=(\tilde{a}_{\mathrm{in},1},\cdots)^{\mathrm{T}}$ and $S$ is the coupling matrix ($S_{j,j^{\prime}}=t_{\mathrm{min}(j,j^{\prime})}\delta_{j,j^{\prime}\pm1}$). The coupling matrix is Hermitian now and can be diagonalized as
\begin{equation}
	S=PJP^{-1}.
\end{equation}
$P=(\bm{\alpha}_{1},\bm{\alpha}_{2},\cdots)$ and the column vectors $\alpha_{j}$ are the eigenvectors of $S$. The diagonal elements of the diagonal matrix $J=\mathrm{Diag}({\lambda_{1},\lambda_{2},\cdots})$ is the corresponding eigenvalues. Then we can rewrite the quantum Langevin equations as
\begin{equation}
	\frac{d}{dt}(P^{-1}\tilde{\bm{\mathrm{A}}})=(-\frac{\kappa}{2}\mathbbm{1}-iJ)P^{-1}\tilde{\bm{\mathrm{A}}}-\sqrt{\kappa}P^{-1}\tilde{\bm{\mathrm{A}}}_{\mathrm{in}}.
\end{equation}
The time-dependent solution is
\begin{equation}
	P^{-1}\tilde{\bm{\mathrm{A}}}(t)=P^{-1}\tilde{\bm{\mathrm{A}}}(0)e^{(-\frac{\kappa}{2}\mathbbm{1}-iJ)t}+\sqrt{\kappa}\int_{0}^{t}e^{(-\frac{\kappa}{2}\mathbbm{1}-iJ)(t-t^{\prime})}P^{-1}\tilde{\bm{\mathrm{A}}}_{\mathrm{in}}(t^{\prime})dt^{\prime}.
\end{equation}
The stationary solution is
\begin{equation}\label{eq:as}
	\begin{split}
		\tilde{\bm{\mathrm{A}}}_{\mathrm{s}}^{\prime}=&\sqrt{\kappa}P\lim_{t\to\infty}\int_{0}^{t}e^{(-\frac{\kappa}{2}\mathbbm{1}-iJ)(t-t^{\prime})}P^{-1}\tilde{\bm{\mathrm{A}}}_{\mathrm{in}}(t^{\prime})dt^{\prime}\\
		=&\sqrt{\kappa}\sum_{j}\lim_{t\to\infty}\int_{0}^{t}e^{(-\frac{\kappa}{2}-i\lambda_{j})(t-t^{\prime})}(\vec{\alpha}_{j}\cdot \tilde{\bm{\mathrm{A}}}_{\mathrm{in}}(t^{\prime}))\vec{\alpha}_{j}dt^{\prime},
	\end{split}
\end{equation}
or
\begin{equation}
	\tilde{a}_{m,\mathrm{s}}=\sqrt{\kappa}\sum_{j,k}\lim_{t\to\infty}\int_{0}^{t}e^{(-\frac{\kappa}{2}-i\lambda_{j})(t-t^{\prime})}\alpha_{j,k}\alpha_{j,m}\tilde{a}_{\mathrm{in},k}(t^{\prime})dt^{\prime}.
\end{equation}
The noise operators before the squeezing transformation satisfy
\begin{equation}
	\langle a_{\mathrm{in},j}(t)a_{\mathrm{in},j}^{\dag}(t^{\prime})\rangle=(n_{\mathrm{th}}+1)\delta(t-t^{\prime}),
\end{equation}
\begin{equation}
	\langle a_{\mathrm{in},j}^{\dag}(t)a_{\mathrm{in},j}(t^{\prime})\rangle=\delta(t-t^{\prime}),
\end{equation}
and the noise operators after the squeezing transformation satisfy
\begin{gather}
	\langle \tilde{a}_{\mathrm{in},j}(t)\tilde{a}_{\mathrm{in},j}^{\dag}(t^{\prime})\rangle=\delta(t-t^{\prime})\frac{(e^{2r_{j}}+e^{-2r_{j}})(2n_{\mathrm{th}}+1)+2}{4},\\
	\langle \tilde{a}_{\mathrm{in},j}^{\dag}(t)\tilde{a}_{\mathrm{in},j}(t^{\prime})\rangle=\delta(t-t^{\prime})\frac{(e^{2r_{j}}+e^{-2r_{j}})(2n_{\mathrm{th}}+1)-2}{4},\\
	\langle \tilde{a}_{\mathrm{in},j}(t)\tilde{a}_{\mathrm{in},j}(t^{\prime})\rangle=\delta(t-t^{\prime})\frac{e^{2r_{j}}-e^{-2r_{j}}}{4}(2n_{\mathrm{th}}+1).
\end{gather}
So the stationary mean values of the second-order moments can be obtained as
\begin{equation}\label{eq:sec1}
	\begin{split}
		\langle \tilde{a}_{m}\tilde{a}_{m^{\prime}}\rangle_{\mathrm{s}}&=\kappa\sum_{j,k,j^{\prime},k^{\prime}}\lim_{t\to\infty}\int_{0}^{t}dt^{\prime}\int_{0}^{t}dt^{\prime\prime}e^{(-\kappa/2-i\lambda_{j})(t-t^{\prime})}\alpha_{j,k}\alpha_{j,m}e^{(-\kappa/2-i\lambda_{j^{\prime}})(t-t^{\prime\prime})}\alpha_{j^{\prime},k^{\prime}}\alpha_{j^{\prime},m^{\prime}}\langle \tilde{a}_{\mathrm{in},k}(t^{\prime})\tilde{a}_{\mathrm{in},k^{\prime}}(t^{\prime\prime})\rangle\\
		&=\kappa\sum_{j,k,j^{\prime}}\lim_{t\to\infty}\int_{0}^{t}dt^{\prime}e^{[-\kappa-i(\lambda_{j}+\lambda_{j^{\prime}})](t-t^{\prime})}\alpha_{j,k}\alpha_{j,m}\alpha_{j^{\prime},k}\alpha_{j^{\prime},m^{\prime}}\frac{e^{2r_{k}}-e^{-2r_{k}}}{4}(2n_{\mathrm{th}}+1)\\
		&=\kappa\sum_{j,k,j^{\prime}}\frac{1}{\kappa+i(\lambda_{j}+\lambda_{j^{\prime}})}\alpha_{j,k}\alpha_{j,m}\alpha_{j^{\prime},k}\alpha_{j^{\prime},m^{\prime}}\frac{e^{2r_{k}}-e^{-2r_{k}}}{4}(2n_{\mathrm{th}}+1),
	\end{split}
\end{equation}
and similarly
\begin{equation}\label{eq:sec2}
	\langle \tilde{a}_{m}^{\dag}\tilde{a}_{m^{\prime}}\rangle_{\mathrm{s}}=\kappa\sum_{j,k,j^{\prime}}\frac{1}{\kappa+i(-\lambda_{j}^{*}+\lambda_{j^{\prime}})}\alpha_{j,k}\alpha_{j,m}\alpha_{j^{\prime},k}\alpha_{j^{\prime},m^{\prime}}\frac{(e^{2r_{k}}+e^{-2r_{k}})(2n_{\mathrm{th}}+1)-2}{4}.
\end{equation}

\end{widetext}

% \bibliography{ref.bib}

	
\end{document}